\documentclass[fleqn,usenatbib]{mnras}

\usepackage{newtxtext,newtxmath}

\usepackage[T1]{fontenc}

\DeclareRobustCommand{\VAN}[3]{#2}
\let\VANthebibliography\thebibliography
\def\thebibliography{\DeclareRobustCommand{\VAN}[3]{##3}\VANthebibliography}

\usepackage{graphicx}	
\usepackage{amsmath}	
\usepackage{pdflscape}

\title[Properties of H\,{\normalsize \textit{II}} regions in barred galaxies]{The impact of bars on the properties of H\,{\Large \textbf{II}} regions in the TIMER survey}

\author[L. S\'anchez-Menguiano et al.]{
Laura S\'anchez-Menguiano,$^{1,2}$\thanks{E-mail: lsanchezm@ugr.es}
Dimitri~A. Gadotti,$^{3}$
Almudena Zurita,$^{1,2}$
Estrella Florido,$^{1,2}$
\newauthor Isabel P\'erez,$^{1,2}$
Paula Coelho,$^{4}$
Jes\'us Falc\'on-Barroso,$^{5,6}$
Taehyun Kim,$^{7}$
Adriana de Lorenzo-C\'aceres,$^{6,5}$
\newauthor Alejandra~Z. Lugo-Aranda,$^{8}$
Justus Neumann,$^{9}$
Camila de S\'a-Freitas$^{10}$
and Patricia S\'anchez-Bl\'azquez$^{11}$
\\
$^{1}$Departamento de F\'isica Te\'orica y del Cosmos, Facultad de Ciencias (Edificio Mecenas), Universidad de Granada, E-18071 Granada, Spain\\
$^{2}$Instituto Carlos I de F\'isica Te\'orica y Computacional, Universidad de Granada, E-18071 Granada, Spain\\
$^{3}$Centre for Extragalactic Astronomy, Department of Physics, Durham University, South Road, Durham DH1 3LE, UK\\
$^{4}$Instituto de Astronomia, Geof\'isica e Ci\^encias Atmosf\'ericas, Universidade de S\~ao Paulo, R. do Mat\~ao 1226, S\~ao Paulo, Brazil\\
$^{5}$Instituto de Astrof\'isica de Canarias, V\'ia L\'actea s/n, 38205 La Laguna, Tenerife, Spain\\
$^{6}$Departamento de Astrof\'isica, Universidad de La Laguna, 38200 La Laguna, Tenerife, Spain\\
$^{7}$Department of Astronomy and Atmospheric Sciences, Kyungpook National University, Daegu 41566, Republic of Korea\\
$^{8}$Universidad Nacional Aut\'onoma de M\'exico, Instituto de Astronom\'ia, AP 106, 22800 Ensenada, BC, M\'exico\\
$^{9}$Max-Planck-Institut f\"ur Astronomie, K\"onigstuhl 17, D-69117 Heidelberg, Germany\\
$^{10}$European Southern Observatory, Alonso de C\'ordova 3107, Vitacura, Regi\'on Metropolitana, Santiago, Chile\\
$^{11}$Departamento de F\'isica de la Tierra y Astrof\'isica, Universidad Complutense de Madrid, E-28040, Madrid, Spain
}

\date{Accepted 2025 November 11. Received 2025 October 31; in original form 2025 February 18}

\pubyear{\the\year{}}

\begin{document}
\label{firstpage}
\pagerange{\pageref{firstpage}--\pageref{lastpage}}
\maketitle

\begin{abstract}
In this study we perform a comparative analysis of the properties of the HII regions located in different areas of barred galaxies, with the aim of investigating the impact of bars on the physical properties of the ionised gas. Based on integral field spectroscopy data for 17 barred galaxies covering approximately the central $6\times6$ kpc, we detect a total of 2200 \ion{H}{ii} regions, of which 331 are located within the nuclear disc (also known as circumnuclear regions),  661 in the bar region, and 1208 in the disc. Among the physical properties of the \ion{H}{ii} regions, we explore the O/H and N/O abundances, H$\alpha$ luminosity, dust extinction, electron density, and H$\alpha$ equivalent width. We find clear differences in the properties of the \ion{H}{ii} regions between the nuclear disc, the bar and the disc, that could be explained by an enhancement in the molecular gas concentration in the central parts driven by bar-induced gas flows. As this gas is channelled towards the galaxy centre, the most extreme values in the analysed properties are found for the circumnuclear \ion{H}{ii} regions. Unlike the bar strength, galaxy mass does seem to affect the properties of the \ion{H}{ii} regions, with massive galaxies presenting higher values in most of the properties, possibly due to the increased amount of gas in these systems. This study provides evidence that the bar-driven redistribution of material within the galaxy inner parts causes significant differences in the \ion{H}{ii} region properties depending on their location within the galaxies.
\end{abstract}




\section{Introduction}

Disc galaxies in the present-day Universe are common hosts of long central stellar structures called bars. It is well established that approximately $30-40\%$ of spiral galaxies exhibit a prominent bar in optical wavelengths, increasing to $60-80\%$ when considering weaker bars, particularly in the near-infrared \citep[e.g.][]{sellwood1993,eskridge2000,menendezdelmestre2007,aguerri2009,masters2011}. These components are proposed as key mechanisms in the dynamical evolution of disc galaxies, contributing to the redistribution of matter by exchanging angular momentum with the disc and dark matter halo and inducing gas flows \citep{lia1992, friedli1994,lia2003}.

The radial motions induced by bars are suggested to mix the gas, altering the abundance distribution within the disc and also in the central region \citep{friedli1994,martin1994,zaritsky1994,portinari2000,cavichia2014, krumholz2017, martel2018}. However, our understanding of the impact of bars on galaxy evolution remains incomplete, particularly in regards to the effects on the properties of the interstellar medium. Evidence suggests that bars enhance the central molecular gas concentration \citep{sakamoto1999,sheth2005, chown2019, querejeta2021, yu2022}, producing a higher star formation rate (SFR) in these inner bar regions \citep{ellison2011, oh2012, catalantorrecilla2017, lin2017, lin2020}. A study by \citet{florido2015} also reports changes in other properties of the ionised gas in the central regions of barred galaxies, namely an increase of dust content, electron density, ionisation parameter, and N/O abundance ratio, the latter likely due to a different star formation history driven by bar-induced gas inflows. A central enhancement in the oxygen abundances has also been reported \citep[e.g.][]{ellison2011}, although this view is not unanimous \citep[e.g.][]{cacho2014, florido2015}. When looking at the radial abundance distribution, early works focused on
investigating the relation of these radial profiles with the properties of their host galaxies report the presence of shallower gradients in barred galaxies compared to unbarred systems \citep[e.g.][]{vilacostas1992, martin1994, zaritsky1994, dutil1999}. However, later studies based on integral field spectroscopy (IFS) with large datasets indicate that bars do not significantly alter the gas metallicity gradient in galaxies \citep{sanchez2014, kaplan2016, sanchezmenguiano2016, sanchezmenguiano2018, wisz2025}. A recent study by \citet{zurita2021b} reveals a flatter abundance gradient when the bar is present, with respect to the one measured in unbarred galaxies, for the lower luminosity (lower mass) galaxies. When analysing the metallicity distribution of spiral arms and the rest of the disc separately, little difference is observed between barred and unbarred galaxies using CALIFA data \citep{sanchezmenguiano2017}. Nonetheless, higher spatial resolution IFS data have shown that bars can significantly enrich the spiral arms compared to the rest of the disc \citep[][]{sanchezmenguiano2016b, ho2017, sanchezmenguiano2020}. All these studies were focused on investigating differences in the integrated properties between barred and unbarred systems. Little is known about the properties of the ionised gas and the \ion{H}{ii} regions within bars, and how much they can differ from those of the ionised gas located in the disc. 

Some works have suggested the idea that star formation (SF) is suppressed in strong bars \citep[e.g.][]{momose2010, hirota2014}, due to the absence of molecular gas (which is quickly funnelled by the bar towards the centre) and/or strong shear preventing the gas to form stars. On the other hand, other studies show that bars are not quiescent structures devoid of molecular gas and that, in fact, higher molecular gas and SFR surface densities are typically found in bars with respect to the galaxy discs \citep{querejeta2021}. There is no uniform picture, however, of where the SF occurs within bars. Regions of SF have often been found near the ends of the bar, in the central regions forming a nuclear ring or circumnuclear starbust, and sometimes also along the bar \citep[e.g.][]{martin1997, verley2007, zurita2008, neumann2019, frasermckelvie2020}. Recently, analysing a large sample of more than 800 galaxies from the {\it Spitzer} Survey of Stellar Structure in Galaxies \citep[S$^4$G;][]{sheth2010}, \citet{diazgarcia2020} studied the distribution of SF in bars, finding that the majority of the galaxies present SF along the bar (class C), a significant fraction exhibit SF at the bar ends, but not along the bar (class B), and a small percentage only have circumnuclear SF (class A). The distribution of SF within the bar is reported to depend on the global physical properties of the host galaxies, with massive, gas-poor, lenticular galaxies typically belonging to SF class A, as a probable consequence of bar-induced SF quenching. Early- and intermediate-type spirals are likely dominated by SF of class B, as a result of higher shear in galaxies with larger central mass concentrations and bar amplitudes. Finally, late-type gas-rich galaxies are mainly assigned to SF class C, probably due to the low shear at the bar region compared to early-type systems. An important aspect to be explored is whether the properties of the ionised gas might also relate to its location within the bar, as the physical conditions that produce the varied observed distributions of SF are clearly different.
 
Recent studies have examined the stellar population properties of bars compared to those of the surrounding discs using IFS data, obtaining more-metal rich and less [Mg/Fe]-enhanced stars in the bars \citep{neumann2020, neumann2024}. This has been proposed to be caused by different mechanisms such as kinematic fractionation during bar formation \citep[trapping metal-rich stellar populations on more elongated orbits, e.g.][]{lia2017, debattista2017,fragkoudi2017}, alongside bar-induced star formation quenching in the inner part of the main galaxy disc, outside the bar but within the bar radius \citep[an area that is usually referred to as star formation deserts, e.g.][]{james2009,james2016}, and mass density-driven efficient enrichment in bars \citep[resulting from the local counterpart to the well-known global stellar mass–stellar metallicity relation, e.g.][]{gonzalezdelgado2014, sanchez2020, neumann2021}. 

In this work, we focus on the properties of the ionised gas, exploring possible differences in the \ion{H}{ii} regions located at different places within the bar, and compared also to those found in the surrounding discs. There is an early study addressing this topic \citep{martin1999} that, although deficient in statistics, was successful in revealing some differences in the population of bar \ion{H}{ii} regions. Based on long-slit spectroscopy, \citet{martin1999} derived the properties of around 50 bar and 40 disc \ion{H}{ii} regions in 10 barred galaxies. With this small sample, the authors found evidence of bar regions, when compared to disc ones, presenting lower ionisation parameters, a similar electron density distribution, higher visual dust extinction, lower H$\alpha$ equivalent width (indicative of older ages, although the authors warn about the large uncertainties introduced by the galactic continuum correction), and a very homogeneous oxygen abundance distribution (probably the result of the large mixing induced by the gas radial flows in bars). More recently, \citet{briere2012} measured the oxygen abundances and stellar population ages of giant \ion{H}{ii} regions in the barred galaxy NGC~5430, finding no variation in both properties between the bar and disc (spiral arms) \ion{H}{ii} regions.

Here we present an updated view on the impact of bars on star formation and the physical properties of \ion{H}{ii} regions, taking advantage of the improvements in spectroscopic techniques, which allow us to increase the statistics in several orders of magnitude. The paper is organised as follows. Section~\ref{sec:sample} briefly describes the data and the galaxy sample, whereas the description of the procedure followed along the analysis is given in Section~\ref{sec:analysis}, including the derivation of gas emission (Section~\ref{sec:gasemission}), the selection and modelling of the \ion{H}{ii} regions (Section~\ref{sec:hiiregions}), the segregation of \ion{H}{ii} regions according to their location within the galaxy (Section~\ref{sec:population}), and the derivation of the analysed properties (Section~\ref{sec:parameters}). The results are presented in Section~\ref{sec:results}, and discussed in Section~\ref{sec:dis}. Finally, Section~\ref{sec:concl} compiles and summarises the main conclusions of the work. 

\begin{table*}
 \caption[]{\label{tab:sample} Main properties of the galaxy sample. These include: galaxy designation (column 1), morphological classification (column 2), galaxy distance (column 3), stellar mass (column 4), position angle and ellipticity of the disc (columns 5 and 6), position angle, ellipticity and semi-major axis of the bar (columns 7-9), bar strength (column 10), and kinematic radius for the nuclear disc (column 11).}
\begin{tabular}{llc@{\hskip 0.7cm}c@{\hskip 0.7cm}c@{\hskip 0.7cm}c@{\hskip 0.8cm}c@{\hskip 0.8cm}c@{\hskip 0.7cm}c@{\hskip 0.7cm}c@{\hskip 0.7cm}c}
 \hline \hline\\[-0.3cm]
 Galaxy & Type$^{\hspace*{0.1cm}a}$ & $d^{\hspace*{0.1cm}b}$ & $M_{\star}$ & PA$_{\rm disc}$ & $\epsilon_{\rm disc}$  & PA$_{\rm bar}$ & $\epsilon_{\rm bar}$ & $R_{\rm bar}$ & $Q_{\rm bar}^{\hspace*{0.3cm}f}$ & $R_{\rm kin}^{\hspace*{0.3cm}g}$\\[0.1cm]
  & & {\footnotesize  [Mpc]} & {\footnotesize  [$\log(M_\odot)$]} & {\footnotesize [deg]} & & {\footnotesize [deg]} & & {\footnotesize [arcsec]} & & {\footnotesize [kpc]}\\[0.2cm]
  (1) & (2) & (3) & (4) & (5) & (6) & (7) & (8) & (9) & (10) & (11) \\ \hline \\[-0.2cm]
IC~1438\hspace{0.9cm}   & SAB$_{\rm a}$0/a\hspace{0.5cm}  & 33.8 & 10.49 & 158$^{\,c}$ & 0.06$^{\,c}$ & 122$^{\,c}$ & 0.54$^{\,c}$ & 26$^{\,c}$ & 0.178 & 0.60 \\
NGC~0613   & SBb & 25.1 & 11.09 & 116$^{\,c}$ & 0.22$^{\,c}$ & 124$^{\,c}$ & 0.70$^{\,c}$ & 80$^{\,c}$ & 0.489 & 0.59 \\
NGC~1097   & SBab pec & 20.0 & 11.24 & 135$^{\,c}$ & 0.31$^{\,c}$ & 148$^{\,c}$ & 0.61$^{\,c}$ & 100$^{\,c}$ & 0.254 & 1.07 \\
NGC~1300   & SBb  & 18.0 & 10.58 & 97$^{\,c}$ & 0.36$^{\,c}$ & 102$^{\,c}$ & 0.77$^{\,c}$ & 77$^{\,c}$ & 0.580 & 0.33 \\
NGC~1365   & SBbc  & 17.9 & 10.98 & 31.66$^{\,d}$ & 0.412$^{\,d}$ & 86$^{\,e}$ & 0.63$^{\,e}$ & 90.4$^{\,e}$ & 0.389 & $-$ \\
NGC~1433   & SBa  & 10.0 & 10.30 & 98$^{\,c}$ & 0.24$^{\,c}$ & 96$^{\,c}$ & 0.71$^{\,c}$ & 74$^{\,c}$ & 0.366 & 0.38 \\
NGC~3351   & SBa & 10.1 & 10.49 & 7$^{\,c}$ & 0.15$^{\,c}$ & 112$^{\,c}$ & 0.57$^{\,c}$ & 56$^{\,c}$ & 0.227 & 0.24 \\
NGC~4303   & SABbc & 16.5 & 10.86 & 72$^{\,c}$ & 0.06$^{\,c}$ & 179$^{\,c}$ & 0.60$^{\,c}$ & 44$^{\,c}$ & 0.535 & 0.21 \\ 
NGC~4981  & SABbc  & 24.7 & 10.45 & 152.36$^{\,d}$ & 0.384$^{\,d}$ & 147$^{\,e}$ & 0.57$^{\,e}$ & 18.9$^{\,e}$ & 0.093 & 0.14 \\ 
NGC~4984   & SAB$_{\rm a}$0/a & 21.3 & 10.60 & 39$^{\,c}$ & 0.11$^{\,c}$ & 91$^{\,c}$ & 0.24$^{\,c}$ & 38$^{\,c}$ & 0.176 & 0.49 \\
NGC~5236   & SABc & 7.0 & 11.04 & 4.08$^{\,d}$ & 0.056$^{\,d}$ & 50$^{\,e}$ & 0.69$^{\,e}$ & 114.9$^{\,e}$ & 0.472 & 0.37 \\
NGC~5248   & SABbc & 16.9 & 10.67 & 103.9$^{\,d}$ & 0.218$^{\,d}$ & 128$^{\,e}$ & 0.36$^{\,e}$ & 27.4$^{\,e}$ & 0.138 & 0.49 \\
NGC~5728   & SB0/a & 30.6 & 10.85 & 32$^{\,c}$ & 0.41$^{\,c}$ & 33$^{\,c}$ & 0.65$^{\,c}$ & 65$^{\,c}$ & 0.387 & 0.63 \\
NGC~6902   & SABab & 38.5 & 10.81 & 56.18$^{\,d}$ & 0.067$^{\,d}$ & 107$^{\,e}$ & 0.19$^{\,e}$ & 10.7$^{\,e}$ & 0.045 & $-$ \\ 
NGC~7140   & SAB$_{\rm x}$ab & 37.4 & 10.71 & 20$^{\,c}$ & 0.42$^{\,c}$ & 20$^{\,c}$ & 0.61$^{\,c}$ & 62$^{\,c}$ & 0.399 & 0.63 \\ 
NGC~7552   & SBa & 17.1 & 10.52 & 115$^{\,c}$ & 0.05$^{\,c}$ & 93$^{\,c}$ & 0.63$^{\,c}$ & 58$^{\,c}$ & 0.358 & 0.33 \\
NGC~7755   & SABbc & 31.5 & 10.60 & 25.49$^{\,d}$ & 0.400$^{\,d}$ & 125$^{\,e}$ & 0.56$^{\,e}$ & 24.6$^{\,e}$ & 0.401 & 0.47 \\ 
\hline
\multicolumn{11}{l}{{\bf References.} (a)~\citet{buta2015}; (b)~\citet{gadotti2019}; (c)~\citet{kim2014};
(d) \citet{salo2015}; (e) \citet{herreraendoqui2015};}\\
\multicolumn{11}{l}{(f) \citet{diazgarcia2016}; (g) \citet{gadotti2020}.}\\
\multicolumn{11}{l}{{\bf Notes.} Information regarding the presence of outer rings or nuclear structures (rings, lenses, bars) has been omitted from the morphological classification}\\
\multicolumn{11}{l}{by \citet{buta2015}. Underline notation was also ignored for being a second-order distinction.}\\
\multicolumn{11}{l}{We have added 90 degrees to the values of PA$_{\rm disc}$ and PA$_{\rm bar}$ extracted from \citet{kim2014} to give both measurements with respect to North.}\\
\end{tabular}
\end{table*}

\section{Galaxy sample}\label{sec:sample}
For this study we make use of the data from the TIMER survey \citep{gadotti2019}. Using the VLT-MUSE integral-field spectrograph \citep{bacon2010, bacon2014}, TIMER has mapped the central part ($6\times6$ kpc on average) of 21 nearby ($d<40$ Mpc) barred galaxies with prominent central structures, such as nuclear rings or nuclear discs, with the main goal of inferring the epoch of bar formation from the star formation histories of this component \citep[see][]{desafreitas2023a, desafreitas2023b, desafreitas2025}. 

In its Wide Field Mode, MUSE presents a field of view (FoV) of approximately $1'\times1'$ and a pixel size of 0.2 arcsec, limiting the spatial resolution of the data to the atmospheric seeing during the observations (which was typically 0.8-0.9 arcsec). At the distances of the galaxies in the sample, this corresponds to approximately between $50-150$ pc. The instrument covers a wavelength range from $\rm 4750$ to $\rm 9300$ \AA, with a spectral sampling of $\rm 1.25$ \AA\ and a spectral resolution $\lambda/\Delta\lambda \sim 1800-3600$ (from the blue edge to the red end of the spectrum). A detailed description of the sample and data properties, as well as observations and data reduction, are given in the survey presentation paper \citep{gadotti2019}.

The spatial distribution of \ion{H}{ii} regions is diverse among the sample: 8 galaxies display extensive SF along the bar region, whereas 9 of them only exhibit localised SF in the central parts\footnote{We note that we do not cover the full bar extension for some galaxies, and therefore, some \ion{H}{ii} regions in these bars might be missing (see Sect.~\ref{sec:caveats} for more details).} or at the ends of the bar. The remaining 4 galaxies do not present any sign of SF in the mapped area and are therefore removed from the sample: these are NGC~1291, NGC~4371, NGC~4643, and NGC~5850. In Table~\ref{tab:sample} we provide the main properties of the galaxy sample. In Fig.~\ref{fig:atlas} (Appendix~\ref{sec:ap1}) we show H$\alpha$ and continuum images of all the galaxies (except for NGC~4981 that can be found in Fig.~\ref{fig:HIIregions}), where the location of SF is easily identified.

\section{Analysis}\label{sec:analysis}

\subsection{Derivation of gas emission}\label{sec:gasemission}
In order to derive the properties of the \ion{H}{ii} regions we first need to determine the emission-line fluxes, which are recovered after a proper subtraction of the stellar population contribution to the observed spectra. This process is done with {\ttfamily PYPARADISE}\footnote{\url{https://github.com/brandherd/PyParadise}}, an extended version of the code {\ttfamily PARADISE} \citep[see][]{walcher2015} developed in {\ttfamily Python}. {\ttfamily PYPARADISE} has been previously used within the TIMER collaboration for the analysis of emission line properties \citep[e.g.][]{gadotti2019, bittner2020, neumann2020}. A detailed description of the procedure can be found in \citet{gadotti2019}, here we provide a brief overview of the main steps. 

First, we estimate the stellar kinematics (stellar velocity and velocity dispersion) following an iterative process by fitting a linear combination of stellar spectra from the template library \citep[we use the Indo-US library;][]{valdes2004} convolved with a Gaussian line-of-sight velocity kernel. For this step, a Voronoi-binning is adopted to increase the signal-to-noise (S/N) of the spectra and obtain an accurate estimation of the stellar kinematics. In our case, a minimum S/N of 40 in the continuum is requested. After that, now in a spaxel-by-spaxel basis, we model the continuum with a linear combination of the stellar template spectra fixing the kinematics according to the underlying Voronoi cell. Finally, the resulting best-fitting stellar spectra are then subtracted from the original ones providing a pure emission line datacube. We then fit the emission lines on the pure gas cube using Gaussian functions. To estimate the uncertainties on all emission-line parameters we repeated the measurements performing 30 Monte Carlo simulations where the input spectra are perturbed within their corresponding variance and afterwards fitted using only 80\% of the template library. This approach enables the propagation of template mismatches in the fitting of the stellar spectra to the uncertainties in the emission-line measurements. The flux, systemic velocity and velocity dispersion (FWHM), together with their corresponding errors, are estimated for all main emission lines, namely H$\beta$, [\ion{O}{iii}]$\lambda\lambda4960, 5007$, [\ion{O}{i}]$\lambda6300$, [\ion{N}{ii}]$\lambda\lambda6548, 6583$, H$\alpha$, and
[\ion{S}{ii}]$\lambda\lambda6717, 6730$.

The  emission-line fluxes are then corrected for dust attenuation following the standard procedure: $F_{\lambda,obs} = F_{\lambda,0} \cdot 10^{-0.4 \, A_\lambda}$, with $F_{\lambda,obs}$ and $F_{\lambda,0}$ the observed and intrinsic emission line flux, respectively, and $A_\lambda$ the absolute extinction at the corresponding wavelength. To determine $A_\lambda$ we make use of the extinction law $A_\lambda/A_V$ from \citet{cardelli1989}, with $R_V=3.1$. The value of $A_V$ is obtained from the H$\alpha$/H$\beta$ Balmer decrement, considering the theoretical value for the unobscured H$\alpha$/H$\beta$ ratio of 2.86, which assumes a case B recombination ($T_e = 10^4$ K, $n_e=10^2$ cm$^{-3}$, \citealt{osterbrock1989}). 

\subsection{Selection and modelling of \ion{H}{ii} regions}\label{sec:hiiregions}
First we detect the ``candidates'' for \ion{H}{ii} regions, that is, clumpy ionised regions in general, using {\ttfamily PYHIIEXTRACTOR} \citep{lugoaranda2022}. This tool is especially designed for high-spatial resolution IFS data like MUSE, where other codes fail to recover the faintest and smallest regions. Written in {\ttfamily Python}, this algorithm brings the novelty of modelling the diffuse ionised gas (DIG), enabling the derivation of the properties of \ion{H}{ii} regions once their emission is decontaminated from the DIG.

\begin{figure}
   \centering
   \includegraphics[width=\hsize]{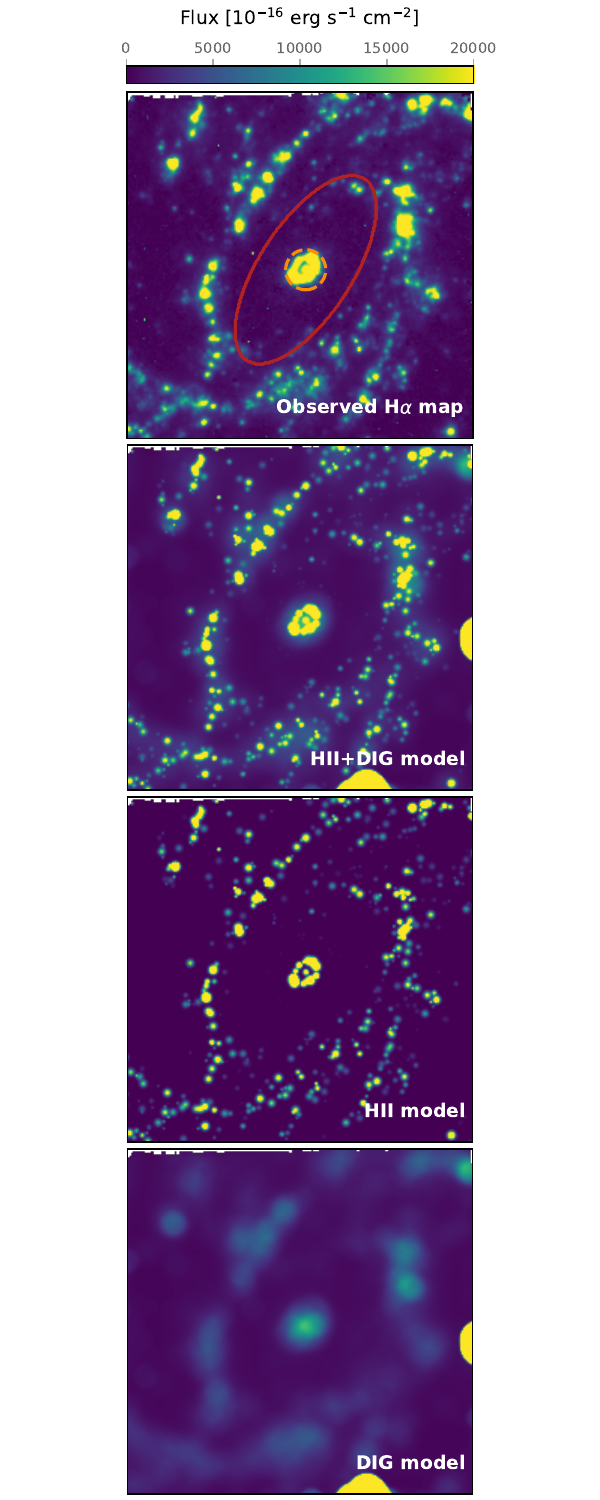}
      \caption{Modelling of the \ion{H}{ii} regions by {\ttfamily PYHIIEXTRACTOR} for the galaxy NGC~4981. We show the observed H$\alpha$ map (top panel), the derived model of the candidate \ion{H}{ii} regions and the DIG together (middle-top panel) and separately (\ion{H}{ii} regions, middle-bottom panel; DIG, bottom panel). Orange circle and red ellipse in top panel delimit the nuclear disc and bar regions, respectively.}
         \label{fig:HIIregions}
   \end{figure}

A detailed description of {\ttfamily PYHIIEXTRACTOR} is provided in \citet{lugoaranda2022}, including some tests on the performance of the algorithm using simulated data with mock galaxies and a comparison with other tools available in the literature developed for similar purposes. Here we summarise the main steps of the process undergone to identify the candidate \ion{H}{ii} regions and extract their main features (positions, radii, fluxes). Firstly, the code identifies the candidates as local maxima in an H$\alpha$ emission map of the galaxy using a blob detection function based on the Laplacian of Gaussian algorithm. For these candidates we obtain the positions and radii. Then, for each provided emission line map, {\ttfamily PYHIIEXTRACTOR} generates a model of these regions assuming a 2D Gaussian distribution for the light profile of each candidate , using the positions and radii obtained with the H$\alpha$ map. By employing fixed values of these parameters for all emission lines we ensure that we are always comparing the emission from exactly the same regions. We note that by using symmetric Gaussian profiles for the modelling of the \ion{H}{ii} regions we are assuming that they are perfectly round regions, which is a valid first-order approximation given that the physical resolutions of the data ($50-150$ pc) are larger than typical \ion{H}{ii} region sizes \citep[e.g.][]{azimlu2011}. This model is subtracted from the emission map, producing a residual map. Assuming that the residual emission should come from the underlying DIG, the code fits this map using a Delaunay tessellation algorithm to find the farthest points from \ion{H}{ii} regions and a kernel interpolation of this residual emission. Finally, this DIG model is subtracted  from the \ion{H}{ii} region model to obtain a clean emission map of the \ion{H}{ii}, regions decontaminated from DIG. From this model the code will derive the flux of each candidate \ion{H}{ii} region integrating the mentioned Gaussian profile. In Figure~\ref{fig:HIIregions} we show an example case of the performance of {\ttfamily PYHIIEXTRACTOR} by detecting and modelling the \ion{H}{ii} regions of the galaxy NGC~4981. We can see that the derived model of the candidate \ion{H}{ii} regions + DIG (middle-top panel) follows very well the light distribution of the observed H$\alpha$ map (top panel), although some artefacts appear at the edges of the image that come from the DIG model itself (bottom panel). This does not affect the analysis since we are only interested in the properties of the \ion{H}{ii} regions (middle-bottom panel), and we have ensured that no \ion{H}{ii} regions fall within the areas of the map affected by these artefacts. 

Candidate \ion{H}{ii} regions presenting $S/N < 3$ in H$\alpha$ emission (estimated as the ratio of the flux to the flux error) are rejected. Lastly, a significant percentage of the sample \citep[around $30\%$, but it depends on the used diagnostic diagram,][]{gadotti2019} is reported to host active galactic nuclei (AGNs) in their centre. It is therefore important to properly select true \ion{H}{ii} regions which emission is dominated by star formation. For that we use the diagnostic WHaD diagram proposed by \citet{sanchez2024}, based on the equivalent width and the velocity dispersion from H$\alpha$. This diagram has the advantage of using information of a high-intensity emission line to classify the ionising sources, thus avoiding weaker lines such as $[\ion{O}{iii}]\lambda5007$ or H$\beta$, whose inclusion can increase the uncertainty in the selection of the \ion{H}{ii} regions. In addition, it has proven more efficient in the identification of AGNs than the traditional BPT diagram \citep{baldwin1981}. In our case, in galaxies such as NGC~1097 or NGC~1365, well-known AGN hosts, we confirm that the WHaD diagram is more restrictive in the selection of \ion{H}{ii} regions, excluding candidates that the traditional BPT diagram classifies as SF. Based on the WHaD diagram, we consider SF \ion{H}{ii} regions the sources with EW(H$\alpha$) > 6 \AA\ and $\sigma$(H$\alpha$) < 57 km s$^{-1}$. On average we detect $\sim 70$ \ion{H}{ii} regions per galaxy, and a total of $2200$ regions in the full sample, discarding around 600 candidates for being predominantly ionised by other sources. A catalogue of strong line fluxes and analysed properties for the resulting \ion{H}{ii} regions is published with this paper. Details on the information included in the catalogue are provided in Appendix~\ref{sec:ap3}.

\subsection{Dissection of \ion{H}{ii} region populations}\label{sec:population}
According to their location within the galaxies, the 2200 \ion{H}{ii} regions have been classified in three groups: 
\begin{enumerate}
\item \ion{H}{ii} regions located inside the nuclear discs\footnote{A nuclear disc is a central stellar structure with an exponential surface density profile built via bar-driven processes that develop in the main galaxy disc \citep{gadotti2020}.}. They are usually distributed in a nuclear star-forming ring, and are also called circumnuclear \ion{H}{ii} regions or hotspots.
\item \ion{H}{ii} regions within the bar, sometimes extended across the bar and sometimes found only at its ends.
\item \ion{H}{ii} regions placed in the main galaxy disc. 
\end{enumerate}

\begin{figure}
   \centering
   \includegraphics[width=\hsize]{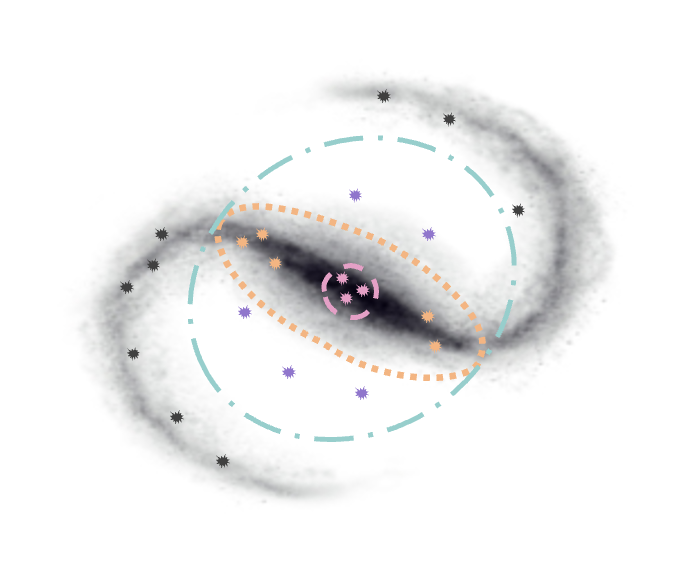}
      \caption{Sketch of a barred galaxy where the different populations of \ion{H}{ii} regions are identified (circumnuclear, pink; bar, orange; inner disc, purple; outer disc, grey). The three contours correspond to the limits of the nuclear disc (dashed pink circle), bar region (dotted orange ellipse), and inner disc (dashed-dotted green ellipse), respectively.
              }
         \label{fig:cartoon}
   \end{figure}
   
The galactocentric distances of the \ion{H}{ii} regions are deprojected using the position angle (PA) and ellipticity ($\epsilon$) of the disc derived by \citet{kim2014}, and by \citet{salo2015} for those galaxies not analysed in the former work. These values are compiled in Table~\ref{tab:sample}. \ion{H}{ii} regions with galactocentric distances lower than $1.5\,R_{\rm kin}$ are classified as circumnuclear, with $R_{\rm kin}$ being the kinematic radius for the nuclear disc. These values are extracted from \citet{gadotti2020} and are available in Table~\ref{tab:sample}\footnote{NGC~1365 and NGC~6902 do not show clear kinematic signatures of the presence of nuclear discs and therefore, $R_{\rm kin}$ is not available in \citet{gadotti2020}. For these galaxies, the central \ion{H}{ii} regions are considered to belong to the bar area.}. We have visually checked that all circumnuclear regions fall within the nuclear disc using this definition and this is the case except for NGC~4981. As these regions are clearly identified as circumnuclear (they form a nuclear star-forming ring, see Fig.{\ref{fig:HIIregions}}) we have extended for this case the nuclear disc border to $3\,R_{\rm kin}$ to include these \ion{H}{ii} regions in the circumnuclear population. We note that we do this for purity reasons (in order to have subsamples of \ion{H}{ii} regions as clean as possible), and that the assignment of these 11 \ion{H}{ii} regions to the circumnuclear or bar populations does not affect the results, since these are statistically robust enough. Dashed orange circles in Fig. \ref{fig:HIIregions} and \ref{fig:atlas} (Appendix~\ref{sec:ap1}) delimit the nuclear disc region in the galaxy sample. 

To identify the \ion{H}{ii} regions within the bar we create masks using PA$_{\rm bar}$, $\epsilon_{\rm bar}$, and bar length ($R_{\rm bar}$) from \citet{kim2014}. For the galaxies not analysed in the former work we use the data from \citet{herreraendoqui2015}. These values are shown in Table~\ref{tab:sample}. The central area corresponding to the nuclear disc is removed from the mask. Red ellipses in Fig. \ref{fig:HIIregions} and \ref{fig:atlas} (Appendix~\ref{sec:ap1}) delimit the bar region in the galaxy sample.

Finally, \ion{H}{ii} regions that are not located within the nuclear disc or the bar are thus considered to belong to the galaxy disc. Simulations have shown that as bars evolve disc stars are captured onto bar orbits \citep{lia2003}. These stars, that are removed from the disc, can make the inner part of the disc surrounding the bar (i.e. at $r < R_{\rm bar}$) become less dense. This light deficit has been observed often in barred galaxies, and has been referred as `dark gaps' \citep{gadotti2003, kim2016, buta2017, ghosh2024}. In order to investigate if these dark gaps affect the properties of the ionised gas, the disc population of \ion{H}{ii} regions is, in turn, split in two subgroups depending on whether they are located inside or outside the bar radius. For that we use their deprojected galactocentric distances, derived as defined above. From now on, we will refer to them as inner and outer disc \ion{H}{ii} regions, respectively.  

To help with the visualisation of the classification, in Fig.~\ref{fig:cartoon} we show a schematic drawing of a barred galaxy where we identify the defined populations of \ion{H}{ii} regions: circumnuclear regions are represented in pink, bar regions in orange, inner disc regions in purple, and outer disc regions in grey. Applying this classification to the entire sample, we end up with 331 circumnuclear \ion{H}{ii} regions, 661 regions belonging to the bar area of the galaxies, and 1208 regions distributed within the discs (of which 260 are located inside the bar radius and 948 outside it). In Table~\ref{tab:sizes1} (Appendix~\ref{sec:ap2}) we break down these numbers for each galaxy in the sample.

\subsection{Derivation of parameters involved in the analysis}\label{sec:parameters}
\subsubsection{O/H and N/O abundances}
Among the analysed parameters, we include both the O/H and N/O abundances of the \ion{H}{ii} regions. Of all metals (understood as elements heavier than helium), oxygen is the most abundant one, constituting nearly 50\% of the metals by number. Together with nitrogen, they produce the strongest emission lines generated by metals and observed in the optical spectral range (due to collisional excitations). For this reason, the oxygen abundance has been considered the best proxy to study the gas metallicity. On the other hand, due to the difference in the mass of the stars producing these two metals (N is ejected to the gaseous interstellar medium mainly by low- and intermediate- mass stars, while O is ejected by massive stars), investigating the N/O ratio could give important clues to the study of the star formation history of the galaxies \citep{molla2006, mallery2007}.

The $T_e$-method, based on the use of temperature-sensitive line ratios, is currently the most direct and reliable technique for measuring chemical abundances from observed spectra \citep[e.g.][]{hagele2006, berg2015, zurita2021a, esteban2025, khoram2025}. However, the faintness of some of these lines, which becomes more pronounced as metallicity increases, hinders their detection and limits its use to very nearby galaxies with high S/N spectra. Thus, calibrators relying on the relationship between metallicity and the intensity of a series of strong emission lines have been proposed to determine abundances in galaxies. These calibrators can be empirical, based on direct abundance estimations, or derived from photoionisation models, with reported offsets of $0.2-0.4$ dex in the results between them \citep[see][for an extended discussion on this issue]{lopezsanchez2012,kewley2019,zurita2021a}.

In this study, although the distance of the galaxies and the S/N of the data might allow us to detect auroral lines in a reduced number of \ion{H}{ii} regions, in order to increase the statistics and the coverage of both the bar and the disc regions, we decide to rely on the use of strong-line estimations of the abundances. The limited wavelength range in the blue regime covered by MUSE, that leaves lines such as [\ion{O}{ii}]$\lambda$3727 undetectable, reduces the number of calibrators available for determining the abundances. For the oxygen abundance, we adopt the empirical S-calibration (Scal) proposed by \citet{pilyugin2016}, which is based on three standard diagnostic line rations:  $N_2 \;\mbox{---}\,I([\ion{N}{ii}]\lambda6548+\lambda6584)/I(H\beta)\,\mbox{---}$, $S_2 \; \mbox{---}\,I([\ion{S}{ii}]\lambda6717+\lambda6731)/I(H\beta)\,\mbox{---}$, and $R_3 \;\mbox{---}\,I([\ion{O}{iii}]\lambda4959+\lambda5007)/I(H\beta)\,\mbox{---}$. To reduce the uncertainty in the oxygen abundance determination, we measure only the strongest line in both the [\ion{O}{iii}] and [\ion{N}{ii}] doublets and assume a fixed ratio of 3:1. The combined use of three emission line ratios makes Scal less dependent on the ionisation parameter \citep{pilyugin2016}. In addition, it has been reported to perform better than other calibrators based on N2 and O3N2 indices \citep[e.g.][]{metha2021}, and similar to some machine-learning approaches \citep{ho2019}. For the N/O abundance, we adopt the empirical calibration proposed by \citet{florido2022} for the N2S2 index (based on the $[\ion{N}{ii}]\lambda6583/[\ion{S}{ii}]\lambda\lambda6717,6731$ line ratio), which relies on a new compilation of 536 \ion{H}{ii} regions with $T_e$-based abundances. Due to the common lines existing in the employed O/H and N/O calibrations, we cannot dismiss the possibility that the O/H estimates are affected by N/O (and vice versa). However, this possible degeneracy is unavoidable due to the limited MUSE wavelength coverage and the unavailability of lines such as [\ion{O}{ii}]. Although Scal uses lines in common with those employed in N2S2 for N/O, is not exactly the same, since it also uses $R_3$, so even if there exists certain degree of uncertainty, it should be minor.

\subsubsection{Dust extinction}
We characterise the dust extinction in the \ion{H}{ii} regions (along the line of sight) using the extinction coefficient in V-band, $A_V$. As described in Sec.~\ref{sec:gasemission}, we use the extinction law from \citet{cardelli1989}, with $R_V=3.1$, and the H$\alpha$/H$\beta$ Balmer decrement, considering the theoretical value for the unobscured H$\alpha$/H$\beta$ ratio of 2.86:
\begin{equation}
A_V \, [{\rm mag}] = 7.217 \times \log_{10} \left( F_{\rm H\alpha}/F_{\rm H\beta} \right) - 3.294
\end{equation}
where $F_{\rm H\alpha}/F_{\rm H\beta}$ is the observed H$\alpha$/H$\beta$ flux ratio. 

\subsubsection{H$\alpha$ luminosity}
The dust-corrected H$\alpha$ luminosity of the \ion{H}{ii} regions, $L({\rm H}\alpha)$, is another analysed parameter. These luminosities are obtained from the integrated dust-corrected H$\alpha$ fluxes of the \ion{H}{ii} regions knowing the galaxy distance $d$ (given in Table~\ref{tab:sample}):
\begin{equation}
L({\rm H}\alpha) \, [{\rm erg \, s}^{-1}] = 4\pi d^2 F({\rm H}\alpha)_{\rm corr}
\end{equation}

\citet{scheuermann2023} recently reported that the $L({\rm H}\alpha)$ of the \ion{H}{ii} regions correlates very strongly with the mass of their ionising stellar associations, with little scatter dependent on the age of the stellar association. We explore how differences in the distribution of $L({\rm H}\alpha)$ values of the \ion{H}{ii} regions between different environments (circumnuclear region, bar, and disc) might be interpreted in terms of cluster mass in those surroundings.

\subsubsection{Stellar mass density}
The stellar surface mass density ($\Sigma_\star$) underlying to the \ion{H}{ii} regions has been derived based on the stellar mass density maps of the galaxies produced by \citet{querejeta2015}. Using the mid-infrared 3.6 $\mu$m band images from the {\it Spitzer} Survey of Stellar Structure in Galaxies \citep[S$^4$G;][]{sheth2010}, the authors applied a method based on Independent Component Analysis to separate the dominant light from the old stars in the 3.6 $\mu$m images from the dust emission. After removing this dust emission, they obtained a smooth distribution of essentially old stellar populations that can be converted into stellar mass using the following relationship:
\begin{equation}
1 \,{\rm MJy/sr} = 704.04 \, L_{\odot}/{\rm pc}^2,
\end{equation}
together with a mass-to-light ratio of $M/L$ = 0.6 $M_\odot/L_\odot$ (assuming a Chabrier IMF). This method provides stellar mass values with an accuracy of $\sim0.1$ dex.

We finally correct each spaxel's surface mass density from projection effects multiplying the observed values by a factor $b/a$, where $a$ and $b$ represent the projected semimajor and semiminor axis of the galaxy discs, respectively. For that we use the ellipticities derived by \citet{kim2014}, and by \citet{salo2015} for the missing galaxies (see Table~\ref{tab:sample}), considering $b/a = 1-\epsilon$. The $\Sigma_\star$ values of the \ion{H}{ii} regions are just the averaged values of the spaxels contained within the regions. 

\subsubsection{Electron density}
The electron density ($N_e$) can be estimated from a sensitive emission line ratio (such as $[\ion{O}{ii}]\lambda3729/[\ion{O}{ii}]\lambda3726$ or $[\ion{S}{ii}]\lambda6717/[\ion{S}{ii}]\lambda6731$), in which the corresponding transitions of the two collisionally excited lines involved, come from very closely spaced energy levels \citep{osterbrock1989}. In our case we directly use the $[\ion{S}{ii}]\lambda6717/[\ion{S}{ii}]\lambda6731$ line ratio\footnote{Typically, in the optical range, with medium resolution spectrographs, only the [\ion{S}{ii}] doublet is  spectrally resolved and sufficiently bright to be useful as a diagnostic tool.} as a proxy for $N_e$. For illustrative purposes, a value of $[\ion{S}{ii}]\lambda6717/[\ion{S}{ii}]\lambda6731 = 1.46$ would correspond to $N_e \sim 1 \,{\rm cm}^{-3}$ (the low-density theoretical limit), and a value of $[\ion{S}{ii}]\lambda6717/[\ion{S}{ii}]\lambda6731 \approx 1.2$ to $N_e \sim 120-200 \,{\rm cm}^{-3}$ for a representative electron temperature of the \ion{H}{ii} region of 8000 K \citep{kewley2019}. When analysing this parameter, we only include \ion{H}{ii} regions whose $[\ion{S}{ii}]\lambda6717/[\ion{S}{ii}]\lambda6731$ line ratio values are below 1.46 (corresponding to densities above the low-density theoretical limit), taking into account their associated uncertainties. However, in the case of the regions whose values are above 1.46 but compatible with this number when considering the uncertainties, due to the unphysical nature of the derived ratios, these are replaced by the low-density limit of 1.46. The reported differences between the distributions when comparing two populations of \ion{H}{ii} regions, if existing, might be slightly overestimated, but very close to the real value since all the replaced $[\ion{S}{ii}]\lambda6717/[\ion{S}{ii}]\lambda6731$ line ratio values should indeed be very close to this number \citep[see section 3.2. and figure 5 of][for details]{barnes2021}.

\subsubsection{H$\alpha$ equivalent width}\label{sec:EW_Ha}
The equivalent width of H$\alpha$ (EW$_{\rm H\alpha}$) measures the relative amount of ionizing and continuum photons emitted by the stars associated with an \ion{H}{ii} region. It depends on the evolutionary status of the stars, the IMF, and the metallicity of the \ion{H}{ii} region. However, evolutionary-synthesis models reveal that the stronger dependence of EW$_{\rm H\alpha}$ comes from the age of the \ion{H}{ii} region \citep[e.g.][]{leitherer1999}. This has been confirmed observationally by means of a clear correlation between log(EW$_{\rm H\alpha}$) and the relative contribution of young stars to the total luminosity of the \ion{H}{ii} regions \citep[e.g.][]{sanchez2014}. Therefore, this parameter can be used as a tracer of the age of an \ion{H}{ii} region, with higher values of EW(H$\alpha$) indicating younger embedded stellar populations.

We derive the equivalent width of the H$\alpha$ emission line for the \ion{H}{ii} regions as 
\begin{equation}
{\rm EW_{H\alpha}} = \frac{F_{\rm H\alpha}}{F_{\rm cH\alpha}} w,
\end{equation}
where $F_{\rm H\alpha}$ is the flux in H$\alpha$ from the \ion{H}{ii} region and $F_{\rm cH\alpha}$ is the flux of the continuum from the underlying stars measured in a window of width $w$ (we choose 60 \AA) adjacent to H$\alpha$. 

We note that, with this definition, within the integration aperture of the \ion{H}{ii} regions we are measuring the contribution from both the ionizing cluster and the underlying population of the disc/bar of the galaxy. The intrinsic difficulties in disentangling the ionizing photons of the \ion{H}{ii} regions from the underlying stars hampers a clean determination of EW$_{\rm H\alpha}$. The implications of this caveat in the use of this parameter as a proxy of the age of the \ion{H}{ii} regions are discussed in Sec.~\ref{sec:dis}. However, \citet{scheuermann2023} show that raw EW$_{\rm H\alpha}$ values strongly correlate with the background corrected ones, and that they also present a robust trend with the \ion{H}{ii} region ages (with a scatter around 10\%, see their figures~5 and 6), which encourages us to include this parameter among the analysed ones despite the existing biases.

\section[Properties of HII regions]{Properties of H\,{\sevensize II} regions}\label{sec:results}

\begin{figure*}
   \centering
   \includegraphics[width=\hsize]{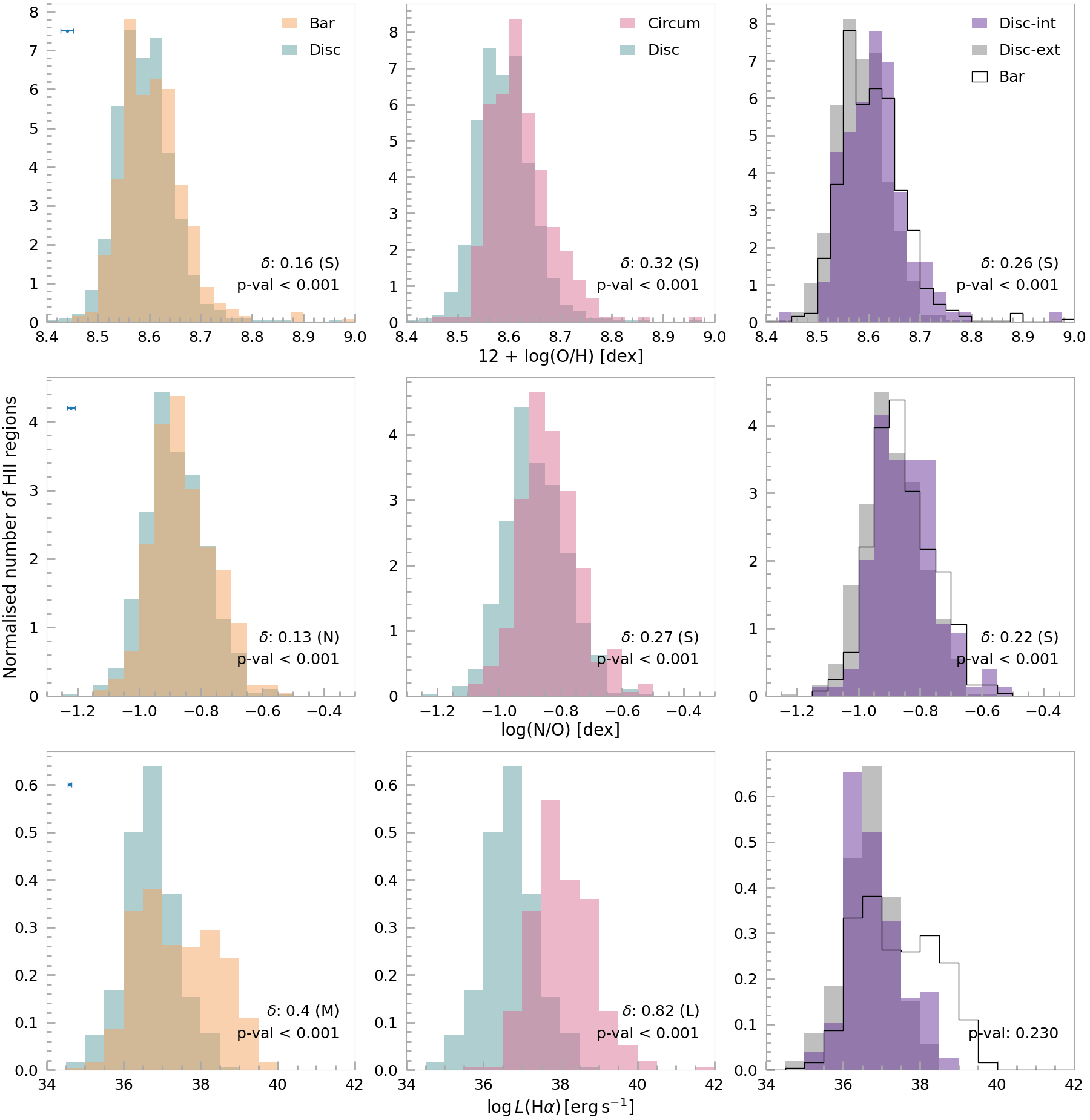}
      \caption{Comparison of the distributions of oxygen abundances ({\it top}), N/O abundances ({\it middle}), and L(H$\alpha$) ({\it bottom}) for different populations of \ion{H}{ii} regions: disc (green) versus bar (orange) regions ({\it left}), disc (green) versus circumnuclear (pink) regions ({\it middle}), and regions in the disc inside (purple) versus outside (grey) the bar radius ({\it right}). The distribution for the bar regions is also represented in right panels with unfilled black histograms. In the top left corner of first panels we include the typical (median) errorbar of each parameter, derived by propagating the errors in the involved emission line intensities (the intrinsic error in the used calibrations is not included for O/H and N/O abundances). In the bottom right corner of the panels the Mann-Whitney U test p-value for each pair of compared distributions is shown, and for those which have values below 5\% (i.e. presenting statistically significant differences), the Cliff's $\delta$ is also provided as a measure of the magnitude of the differences. These are categorised as negligible (N), small (S), medium (M) and large (L) differences (see Sec.~\ref{sec:results} for details on these statistics).}
         \label{fig:hists1}
   \end{figure*}
   
We explore the existence of differences in the properties of the \ion{H}{ii} regions depending on their location within each barred galaxy in our sample. For that we compare the distributions of the analysed parameters for the different \ion{H}{ii} region populations described in Sec.~\ref{sec:population}.

In order to quantify the similarity/difference of the distributions we use two statistical tests. The first one, the Mann-Whitney U test, is a non-parametric test that analyses whether or not two samples come from the same distribution. It is the non-parametric version of the t-test for independent samples, the latter being only valid for normally distributed populations. With this test, a p-value lower than 5\% indicates that we can reject the null hypothesis in favour of the alternative that the distributions are different. However, this statistical hypothesis tests the likelihood that two distributions are different, but it does not give information on the magnitude of the differences. The differences could be significant, but the magnitude so small that it has little impact. Therefore, in order to complement the U test, in the cases where it yields differences in the parameter distribution of two \ion{H}{ii} region populations, we use the Cliff's delta ($\delta$) measure to provide an estimate of the magnitude of the differences. This non-parametric statistic measures the non-overlapping area of two distributions, and does not make any assumptions about the underlying distributional properties, being appropriate for non-Gaussian distributed data. The $\delta$ parameter can be interpreted as the degree of non-overlap between the two distributions. For instance, $|\delta|=0.4$ indicates a 40\% non-overlap (or 60\% overlap) between both distributions. It is usual to classify effect sizes as ``negligible'', ``small'', ``medium'', or ``large''. A $|\delta|$ value lower than 0.15 indicates negligible differences, between 0.15 and 0.33 small differences, between 0.33 and 0.47 medium differences, and if $\delta$ is higher than 0.47 means that the differences are large \citep{meissel2024}. We note that $\delta$ can present positive and negative values, the latter just meaning the reference distribution (one of the compared two) presents overall lower values of the analysed property than the other distribution. If the reference distribution shows larger values than the comparing distribution, then $\delta>0$. In our case, the reference distribution is the one of bar \ion{H}{ii} regions in the bar-disc comparison, circumnuclear \ion{H}{ii} regions in the circumnuclear-disc comparison, and inner disc \ion{H}{ii} regions in the inner-outer disc comparison.

\begin{figure*}
   \centering
   \includegraphics[width=\hsize]{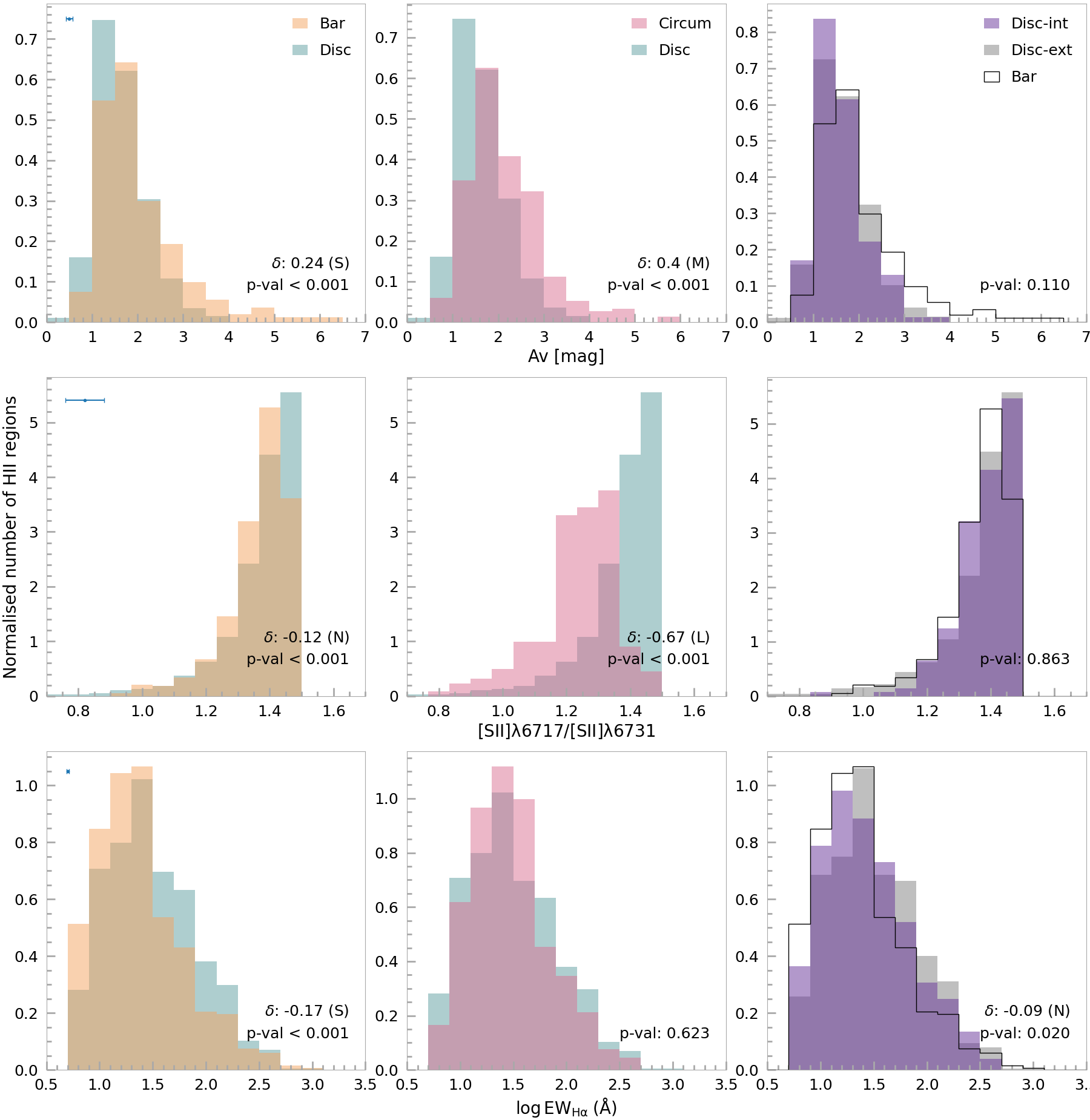}
      \caption{Same as Fig.~\ref{fig:hists1} but for the dust extinction ({\it top}), the $[\ion{S}{ii}]\lambda6717/[\ion{S}{ii}]\lambda6731$ line ratio as a proxy for the electron density ({\it middle}), and the H$\alpha$ equivalent width ({\it bottom}). See caption above for more details.}
         \label{fig:hists2}
   \end{figure*}
   
\subsection{Chemical abundances}\label{sec:abun}
The top row of Fig.~\ref{fig:hists1} displays the histograms of oxygen abundance values for all the \ion{H}{ii} region populations defined in Sec.~\ref{sec:population}. We compare the distributions of the \ion{H}{ii} regions found in the disc (green) and the bar (orange) in the left panel, the disc and the circumnuclear (pink) \ion{H}{ii} regions in the middle panel, and the regions located in the disc inside (purple) and outside (grey) the bar radius in the right panel. We can see that the circumnuclear and bar regions show O/H values that are slightly higher than those of the disc regions, and within the disc, the inner regions are also slightly more metal-rich than the outer ones. The p-value of the Mann-Whitney U test yields statistically significant differences between the compared distributions in the three cases (p-value $<0.1\%$), although the Cliff's $\delta$ indicates that these differences are small ($\delta$ < 0.33). The higher O/H values are found for the circumnuclear \ion{H}{ii} regions, whose distribution present a non-overlap of 32\% with the distribution of the disc regions, the largest differences found in the comparison.

We also compare in the middle row of Fig.~\ref{fig:hists1} the distribution of N/O values for the same subsets of \ion{H}{ii} regions, with very similar trends to those observed for the oxygen abundances: on average, circumnuclear and bar regions present higher N/O abundances compared to disc \ion{H}{ii} regions, as well as disc regions inside the bar radius compared to those located outside it. Again, in all cases the Mann-Whitney U test shows that the distributions are statistically different (p-value $<0.1\%$), but the magnitude of the differences is small, with Cliff's $\delta$ ranging between 0.13 and 0.27 (13-27\% of non-overlap between the distributions).

In Sec.~\ref{sec:dis} we discuss if these differences only account for the well-known negative metallicity gradients observed in spiral galaxies (\citealt{searle1971}; see \citealt{maiolino2019} and \citealt{sanchez2020} for a review on the topic), result of their inside-out growth \citep{matteucci1989, boissier1999}, due to the different ranges of galactocentric distances covered by each population of \ion{H}{ii} regions. It is relevant to assess if, besides this, the bar plays an additional role in the chemical enrichment of the \ion{H}{ii} regions influenced by its action.

\subsection{H$\alpha$ luminosity}
In the bottom row of Fig.~\ref{fig:hists1} we show the distribution of H$\alpha$ luminosities for the different populations of \ion{H}{ii} regions. In this case it is clear that both bar and circumnuclear \ion{H}{ii} regions are more luminous than the disc \ion{H}{ii} regions, with Cliff's $\delta$ values of 0.4 and 0.82, indicating an overlap of 60\% and 18\% with respect to the distribution of the disc regions, respectively. The bar \ion{H}{ii} regions are on average 4 times more luminous than the disc regions, and the circumnuclear, 17 times more luminous. Finally, if we compare the distributions for the disc regions we can see that both subsamples (\ion{H}{ii} regions located inside and outside the bar radius) present very similar distributions, as revealed by the Mann-Whitney U test (p-value of 23\%).

\subsection{Dust extinction}
The top row of Fig.~\ref{fig:hists2} represents the distributions of $A_V$ for the compared subsamples. We find small differences between the \ion{H}{ii} regions in the bar and the disc, supported by a p-value below 0.1\% and a $\delta=0.24$. The dust extinction in the \ion{H}{ii} regions of the bar is on average 0.2 mag higher than in the \ion{H}{ii} regions of the disc. Differences between circumnuclear and disc \ion{H}{ii} regions are even larger, with a $\delta$ value of 0.4. On average, circumnuclear \ion{H}{ii} regions have 0.4 mag higher dust extinction than the disc \ion{H}{ii} regions. Within the disc, however, no differences have been found between the inner and outer \ion{H}{ii} regions, with the p-value of 11\% indicating that the $A_V$ distributions are indeed drawn from the same parent distribution. 

\subsection{Electron density}
In the middle row of Fig.~\ref{fig:hists2} we explore how the electron density of the \ion{H}{ii} regions, estimated by the $[\ion{S}{ii}]\lambda6717/[\ion{S}{ii}]\lambda6731$ line ratio, is affected by their location within the galaxy. The only significant differences found for the electron density appear when comparing the circumnuclear and disc \ion{H}{ii} regions. In this case, \ion{H}{ii} regions in the nuclear disc present lower values of this line ratio compared to the \ion{H}{ii} regions in the disc, which is indicative of higher $N_e$ values. This is supported by a p-value below 0.1\% in the Mann-Whitney U test and a $\delta=-0.67$. On average, the $[\ion{S}{ii}]\lambda6717/[\ion{S}{ii}]\lambda6731$ line ratio is 0.15 lower in circumnuclear \ion{H}{ii} regions than in disc regions. When comparing the distributions between the regions located in the disc and the bar, we obtain a p-value lower than 0.1\%, indicating that the distributions are not drawn from the same parent one. However, the magnitude of the differences is negligible according to the reported $\delta$ value of -0.12. In addition, very similar distributions are found when comparing the inner and outer disc \ion{H}{ii} regions, as demonstrated by the obtained p-value (87\%).

\begin{figure*}
   \centering
   \includegraphics[width=0.7\hsize]{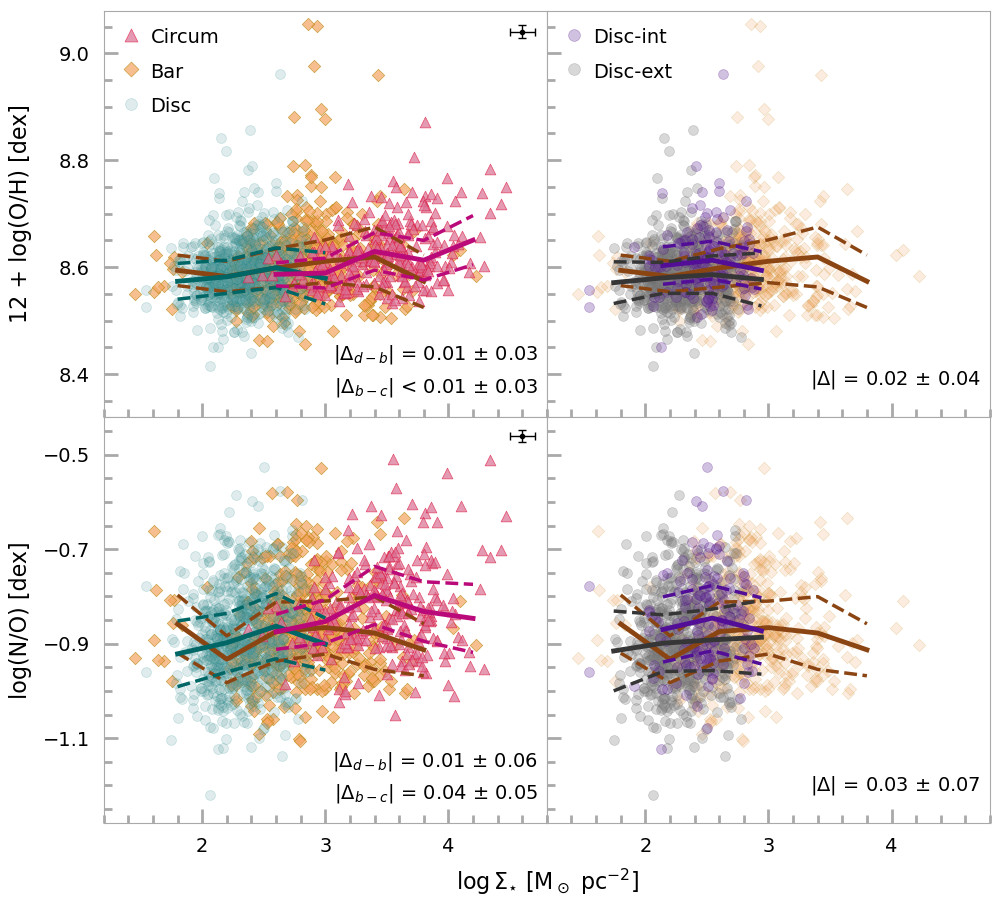}
      \caption{Resolved $\Sigma_\star-{\rm O/H}$ ({\it top}) and $\Sigma_\star-{\rm N/O}$ ({\it bottom}) relations. \ion{H}{ii} regions have been segregated as in Fig.~\ref{fig:hists1}, following Sec.~\ref{sec:population}. Solid and dashed lines mark the median trend and the median absolute deviation, respectively, in bins of 0.4 log $\Sigma_\star$. The average absolute difference of the median trend between the two represented distributions (|$\Delta$|) is given in the bottom right corner of the panels.}
         \label{fig:mzr}
   \end{figure*}
   
\subsection{H$\alpha$ equivalent width}
In the bottom row of Fig.~\ref{fig:hists2} we compare the distributions of EW$_{\rm H\alpha}$ for all the analysed subsamples of \ion{H}{ii} regions. In this case, only small differences are shown when we compare the distributions of \ion{H}{ii} regions in the bar and the disc ($\delta=-0.17$). On average, \ion{H}{ii} regions within the bar have EW$_{\rm H\alpha}$ values that are 6.7 \AA\ lower than those located in the disc. Circumnuclear and disc \ion{H}{ii} regions present similar EW$_{\rm H\alpha}$ distributions, supported by a p-value of 62\%. Finally, although the Mann-Whitney U test suggests that the distributions of EW$_{\rm H\alpha}$ are indeed different for the inner and outer disc \ion{H}{ii} regions (p-value of 2\%), we find that these differences are negligible when deriving the Cliff's $\delta$ ($\delta=-0.09$).

\section{Discussion}\label{sec:dis}
In this paper we perform a comparative analysis of the properties of the \ion{H}{ii} regions located in different areas of barred galaxies, with the main aim of investigating the impact of bars on star formation and the physical properties of the ionised gas. Our analysis is based on IFS data for 17 galaxies from the TIMER survey, with an average spatial resolution of $80-90$ pc. In this study we increase the statistics in various orders of magnitudes with respect to previous works, going from tens to thousands of \ion{H}{ii} regions. 

\subsection{Chemical differences in the \ion{H}{ii} regions: relation with changes in stellar mass surface density}\label{sec:rMZR}
   
In Sec.~\ref{sec:abun}, we show that the \ion{H}{ii} regions located in the bar and the nuclear disc (also referred to as circumnuclear \ion{H}{ii} regions) are slightly more metal rich than the regions in the disc. Within the disc, inner \ion{H}{ii} regions (inside the bar radius) have also higher O/H abundances than the outer regions (outside the bar radius). The significance of the differences in the distributions is confirmed by the Mann-Whitney U tests and the Cliff's $\delta$ values. However, despite this statistically significant dissimilarity in the distributions, the reported differences in the medians are small, of only 0.02-0.03 dex. Similarly, statistically significant higher N/O abundances are found for the bar and circumnuclear \ion{H}{ii} regions compared to the disc, and also within the disc itself between inner and outer regions, supported by the performed tests, but the differences in the medians are again of only 0.02-0.05 dex.

Metals such O and N are a direct by-product of star formation in galaxies, its amounts increasing as successive generations of stars expel them to the interstellar medium (ISM) after their death, modulated by the inflow or outflow of gas in different regions. As a natural consequence of this process, the accumulated stellar mass of the galaxy (and also its local counterpart in the SF areas, $\Sigma_\star$) also increases. These two parameters are therefore connected through the well-known mass-metallicity relation \citep[MZR,][]{tremonti2004}, which can be recovered from its resolved version (a relation between $\Sigma_\star$ and O/H, rMZR hereafter) by integrating the local one along the optical extension of the galaxy \citep[e.g.][]{rosalesortega2012, sanchez2013, barreraballesteros2016}. An analogous relation has also been reported for N/O \citep[rMNOR hereafter,][]{perezmontero2009}. 

Observational studies have shown that bars are associated with stellar overdensities \citep[e.g.][]{kim2016, krishnarao2022, neumann2024}, an intrinsic result of disc stars being trapped on the bar orbits as bars evolve \citep{lia2003}. These higher stellar mass surface densities could be the cause of our finding of more metal rich \ion{H}{ii} regions in the bar compared to the disc. It is thus imperative to clarify whether bar \ion{H}{ii} regions are truly chemically enriched, or our results are just reflecting differences in $\Sigma_\star$ in accordance with the rMZR and the rMNOR. In the top-left panel of Fig.~\ref{fig:mzr} we show the $\Sigma_\star-{\rm O/H}$ relation for circumnuclear (pink), bar (orange), and disc (green) \ion{H}{ii} regions. It is clear that regions in the nuclear disc and the bar reach higher $\Sigma_\star$, covering a wide range of values from $2.4-4.6$ and $1.6-4.2 \rm \log M_\odot \, pc^{-2}$, respectively, while disc \ion{H}{ii} regions are much less dense ($1.6-2.8 \rm \log M_\odot \, pc^{-2}$). However, the three populations of \ion{H}{ii} regions follow the same trend along the rMZR (solid lines), with similar median absolute deviations (dashed lines). In the top-right panel of Fig.~\ref{fig:mzr} we show the same relation for inner (purple) and outer (grey) disc \ion{H}{ii} regions. We can see that the median trend for the inner regions is always above the trend for the outer regions (and generally above the one for the bar regions). At fixed $\Sigma_\star$, inner disc \ion{H}{ii} regions are 0.02-0.03 dex more metal-rich than the outer disc and bar regions. However, the differences are small, and further investigation is needed to confirm these tentative trends.

In the case of the $\Sigma_\star-{\rm N/O}$ relation (bottom-left panel), something equivalent is observed for the comparison of the \ion{H}{ii} regions in the disc and the bar, although with a larger scatter. For the circumnuclear \ion{H}{ii} regions, however, at the highest stellar mass surface densities ($\Sigma_\star > 3.2 \rm \log M_\odot \, pc^{-2}$), we find higher values of N/O ($\sim 0.08$ dex) compared with the \ion{H}{ii} regions in the bar (at these high densities there are no regions in the disc to compare with). The number of \ion{H}{ii} regions is large enough for the result to be statistically significant (64 bar and 177 circumnuclear regions). When comparing the inner and outer disc \ion{H}{ii} regions (bottom-right panel), as for O/H, the median trend of the inner regions is systematically 0.03-0.04 dex above that of the outer regions (and generally above the one for the bar regions), hinting that at fixed $\Sigma_\star$, inner disc \ion{H}{ii} regions are slightly more metal-rich than the outer disc and bar regions, but again this has to be confirmed due to the small magnitude of the differences compared with the scatter.

Thus, we can conclude that the higher $\Sigma_\star$ values found in the bar area result in \ion{H}{ii} regions in the bar having higher O/H and N/O abundances with respect to the disc regions, but at fixed $\Sigma_\star$, bar and disc \ion{H}{ii} regions present a similar degree of chemical enrichment. Within the disc, inner regions are slightly more metal-rich than the outer regions, but this finding is inconclusive and needs further confirmation. Finally, we do find significantly higher N/O values in the circumnuclear \ion{H}{ii} regions compared to the bar regions at similar $\Sigma_\star$ values, for which we give a possible explanation in Sec~\ref{sec:interpretation}. In this section we have interpreted the chemical differences in terms of the rMZR. Alternatively, we could have explored the results in terms of metallicity gradients and differences in the covered radial range between the various populations of \ion{H}{ii} regions. Our strategy has been chosen based on limitations in the radial coverage of the galaxies, but in Sec.~\ref{sec:caveats} we argue that both approaches are equivalent and indeed the differences in the oxygen abundance of the \ion{H}{ii} regions depending on their location within the galaxy are a consequence of the underlying radial negative gradient.

\subsection{Interpreting the observed differences in the \ion{H}{ii} region properties}\label{sec:interpretation}

In Sec.~\ref{sec:results} we report clear differences in the analysed properties of the \ion{H}{ii} regions depending on whether they are located on the nuclear disc, the bar or the main disc. These differences could be explained by an enhancement in the molecular gas concentration in the central parts driven by bar-induced gas flows \citep{sheth2005, querejeta2021}. Since this gas is channelled towards the galaxy centre, the most extreme values in the analysed properties are found for \ion{H}{ii} regions within the nuclear disc (middle panels in Fig.~\ref{fig:hists1} and \ref{fig:hists2}). \ion{H}{ii} regions in the nuclear disc are usually found in the shape of a ring, delimiting the outer rim of the nuclear disc \citep{cole2014}. In barred galaxies, nuclear discs are presumably formed from gas brought to the central region by the bar. As it falls into the outer boundary of the region dominated by the $x_2$ orbits\footnote{$x_2$ is a family of periodic orbits found in the bar inner region with low eccentricity and perpendicular to the bar major axis \citep{buta1996}.}, the gas is prone to form stars and produce star-forming nuclear discs \citep{gadotti2020}. These gas flows induced by the bar enhance the molecular gas concentration in the inner regions \citep{sakamoto1999,sheth2005,chown2019}. This denser gas accumulated in the central parts could explain the presence of more H$\alpha$-luminous \ion{H}{II} regions \citep[associated with more massive ionising star clusters,][]{scheuermann2023} since it has been shown that molecular clouds forming more massive stars tend to have higher average column densities \citep[e.g.][]{schneider2015}. The enhancement in the molecular gas concentration would also produce the observed increase in the dust content (characterised by dust extinction), since dust particles are normally found associated with molecular gas \citep[e.g.][]{lombardi2006}. In addition, the greater concentration of molecular gas and dust in the nuclear disc could also lead to the reported enhancement in the electron densities. Since stars (whose high-energy radiation ionise the surrounding gas to form the \ion{H}{ii} region) are born in the centre of molecular clouds, the initial electron densities of the \ion{H}{ii} region is likely to be set by the molecular gas number density, suggesting indeed a correlation between $N_e$ and molecular gas concentration \citep{davies2021}. However, since the feedback timescales are short, it is likely that the molecular gas in the vicinity of the \ion{H}{II} regions has already been cleared \citep{kim2021, chevance2022}, and therefore, the enhancement in the electron density might be simply due to the high pressure and density ambient that usually characterises the galaxy centres \citep{kruijssen2013, barnes2020, chevance2020}.

Although the largest differences are found for the circumnuclear \ion{H}{ii} regions, it is important to emphasise that bar-disc differences also exist. This is clear when comparing the distributions of the analysed properties for the disc and bar \ion{H}{ii} region populations (left panels in Fig.~\ref{fig:hists1} and \ref{fig:hists2}). Bar \ion{H}{ii} regions present intermediate values in their properties between circumnuclear and disc \ion{H}{ii} regions, which suggests that the bar is a transitional location for the gas until it reaches the innermost regions in the galaxy. During this transportation, the bar would also suffer an enhancement in molecular gas density \citep{querejeta2021}, accompanied by the observed increase of dust content and mass of the star associations ionising the \ion{H}{ii} regions.

The gas inflow induced by the bar can also affect the star formation history in the centre of the galaxies, as it is suggested by the higher N/O abundances found for the circumnuclear \ion{H}{ii} regions \citep[due to the time delay in the ejection of these two metals to the ISM,][]{molla2006, mallery2007}. In particular, an increase in N/O is associated with a decrease in the SFR with time (assuming a closed-box model): if fewer massive stars are being formed in the present, this means that the amount of O that is released to the ISM by these new massive stars will be lower than the amount of N released by intermediate-mass stars from previous generations (when SFR was higher). This scenario holds even in the presence of metal-poor gas flows, due to the low sensitiveness of this ratio to gas infall \citep[large mass infall and infall rate should exist for their effect to be noticeable,][]{koppen2005}. As suggested by \citet{florido2015}, our finding of higher N/O in the circumnuclear \ion{H}{ii} regions could be qualitatively explained according to this picture. Numerical simulations predict that, as the bar evolves, the availability of gas in the disc diminishes, decreasing hence the rate at which the material is transported to the galaxy centre \citep{lia2005}. This implies that when the bar forms, the strong gas flows can lead to an intense SF, that would decline as the bar weakens and the gas flows transport material at a lower rate. 

Lastly, we have also investigated the existence of differences in the properties of inner and outer disc \ion{H}{ii} regions. Overall (and except for the chemical content), it seems that the population of disc \ion{H}{ii} regions present very similar properties irrespectively of whether they are inside or outside the bar radius. However, we must keep in mind that although we make the distinction of inner and outer disc \ion{H}{ii} regions, we are restricting the analysis to the very central part of the galaxies ($\sim6 \times 6$ kpc, on average), and different results might be reached if extending the study to larger galactocentric distances. The slight chemical enrichment found in Sec.~\ref{sec:rMZR} for the inner disc regions with respect to the outer disc and bar regions might reflect some physical mechanism at play behind. Indeed, we know that bars efficiently redistribute material, bringing gas from the outer to the inner parts of the galaxies \citep{lia1992,friedli1994}. This gas, which would be more metal poor due to the well-known negative metallicity gradients displayed by galaxies \citep[e.g.][]{sanchezmenguiano2016, sanchezmenguiano2018, zurita2021b}, is mixed within the bar, which would produce the bar \ion{H}{ii} regions to be more metal-poor than the \ion{H}{ii} regions in the disc at $r < R_{\rm bar}$, since the latter have not interacted with the funnelled metal-poor gas. However, the magnitude of the differences is small compared with the scatter, and further confirmation is needed. Indeed, in \citet{zurita2021b} they measured flatter metallicity radial profiles in barred galaxies compared with unbarred systems even excluding the \ion{H}{ii} regions in the bar, suggesting that the mixing is also observed outside the bar region, which would contradict our scenario in this regard.

\subsection{Effect of global galaxy properties}
We now explore how the properties of the \ion{H}{ii} regions might be influenced by two global galaxy parameters depending on their location within the galaxy. We focus on the bar strength and the galaxy mass. The narrow range of morphological types covered by the sample (mostly Sa-Sbc), prevents us from including this parameter in the analysis. By segregating the galaxies in two subgroups based on each explored parameter, we end up with a very low number of \ion{H}{ii} regions in the inner disc in the case of low-mass and weakly-barred galaxies ($\sim20-40$). Therefore, we avoid this internal division within the disc population and examine differences in the properties of the \ion{H}{ii} regions in the full disc, the bar, and the nuclear disc.

\subsubsection{Bar strength}
Our results suggest that the gas flows induced by the bar are able to modify the properties of the ionised gas. Simulations show that the gas inflow is strongly dependent on the bar strength, with stronger bars supplying gas towards the galaxy centres at a higher rate \citep[e.g.][]{lia1992, regan2004, hopkins2011, kim2012}, although it will also depend on the availability of gas in the disc. Considering both facts together, differences in the properties of the \ion{H}{ii} regions in the bar and the nuclear disc could be expected when comparing galaxies with weak and strong bars. To test this hypothesis we divide the sample in two subgroups according to the strength of their bar. To characterise the bar strength we use $Q_{\rm bar}$, which represents the maximum bar torque applied to a gaseous material in orbital motion relative to its specific kinetic energy. $Q_{\rm bar}$ values for all galaxies in the sample are presented in Table~\ref{tab:sample}. These are obtained in \citet{diazgarcia2016} from the ratio of the tangential to the mean axisymmetric radial force. We consider $Q_{\rm bar}=0.3$ as the limit between strong and weak bars. Other parameters have also been proposed as a numerical representation of the bar strength \citep[see][for a discussion on different methods]{buta2001}, one of the simplest being the deprojected bar ellipticity \citep{martin1995b}. We note that in our sample there is a tight correlation between $Q_{\rm bar}$ and $\epsilon_{\rm bar}$ ($r=0.77$), and reproducing the analysis using $\epsilon_{\rm bar}$ leads to similar results when comparing the distributions of the \ion{H}{ii} region properties. We only find one difference, that will be detailed below.

\begin{table}
\centering
 \caption[]{\label{tab:cliffs} Mann-Whitney U test p-values and Cliff's $\delta$ values for the analysed properties of the compared subpopulations of \ion{H}{ii} regions (circumnuclear, bar, and disc) when segregating the sample according to the galaxy mass and the bar strength.}
\begin{tabular}{l@{\hskip 0.2cm}c@{\hskip 0.7cm}c}
 \hline \hline\\[-0.3cm]
  & {\bf Bar strength} & {\bf Galaxy mass}\\
  & Weak vs. strong bars & Low- vs. high-mass\\[0.1cm]
  & ($Q_{\rm b,lim} = 0.3$) & ($\log M_{\rm \star, \,lim} = 10.65 \,{\rm M_\odot}$) \\ \hline \\[-0.2cm]
{\bf 12+log(O/H)} &  &  \\
Disc & $37.4\%$ & $<0.1\% \;(\delta=-0.22)$ \\
Bar & $56.1\%$ & $<0.1\% \;(\delta=-0.31)$ \\
Circum & $54.7\%$ & $<0.1\% \;(\delta=-0.26)$ \\[0.2cm]
{\bf log(N/O)} &  &   \\
Disc & $15.8\%$ & $<0.1\% \;(\delta=-0.27)$ \\
Bar & $5.1\%$ & $<0.1\% \;(\delta=-0.39)$ \\
Circum & $<0.1\% \;(\delta=0.27)$ & $<0.1\% \;(\delta=-0.43)$ \\[0.2cm]
{\bf $\mathbf{L(H\alpha})$} &  &  \\
Disc & $<0.1\% \;(\delta=-0.36)$ & $<0.1\% \;(\delta=0.27)$ \\
Bar & $64.4\%$ & $<0.1\% \;(\delta=0.65)$ \\
Circum & $39.0\%$ & $15.0\%$ \\[0.2cm]
$\mathbf{A_V}$ &  & \\
Disc & $96.0\%$ & $64.5\%$ \\
Bar & $1.5\% \;(\delta=0.15)$ & $<0.1\% \;(\delta=-0.20)$ \\
Circum & $77.5\%$ & $<0.1\% \;(\delta=-0.32)$ \\[0.2cm]
{\bf $\mathbf{N_e}$} &  & \\
Disc & $68.1\%$ & $<0.1\% \;(\delta=0.08)$  \\
Bar & $61.9\%$ & $<0.1\% \;(\delta=0.19)$  \\
Circum & $6.3\%$ & $<0.1\% \;(\delta=0.22)$ \\[0.2cm]
{\bf $\mathbf{ EW_{H\alpha}}$} &  & \\
Disc & $3.2\% \;(\delta=0.07)$ & $<0.1\% \;(\delta=0.10)$ \\
Bar & $49.2\%$ & $9.7\%$\\
Circum & $16.3\%$ & $1.7\% \;(\delta=-0.15)$ \\
\hline
\multicolumn{3}{l}{We show the Cliff's $\delta$ values in brackets only for the comparison of the}\\
\multicolumn{3}{l}{distributions for which the Mann-Whitney U test yields p-values below}\\
\multicolumn{3}{l}{5\%, indicating that the distributions are not drawn for the same parent}\\
\multicolumn{3}{l}{population.}\\
\end{tabular}
\end{table}

In the second column of Table~\ref{tab:cliffs} we show the Mann-Whitney U test p-values for the analysed properties of the compared subpopulations of \ion{H}{ii} regions when segregating the sample according to the bar strength. That is, we compare the distribution of values for each analysed property and subpopulation of \ion{H}{ii} regions (disc, bar, and circumnuclear) between galaxies with weak and strong bars. We also provide the Cliff's $\delta$ values (in brackets) for the pairs of distributions for which the Mann-Whitney U test yields p-values below 5\%, indicating that both distributions are not drawn for the same parent population. Inspecting the numbers, there is no general trend of finding significant differences in the properties of the \ion{H}{ii} regions for any of the analysed subpopulations that could be induced by the bar strength. The differences found are few and involve only one or two properties. In the case of the \ion{H}{ii} regions located in the disc, only for the $L({\rm H}\alpha)$ we find differences in the distributions supported by both the Mann-Whitney U test and the Cliff's $\delta$ parameter, with \ion{H}{ii} regions being 2.5 times more luminous when a strong bar is present. For the $\rm EW_{H\alpha}$, although existing, the differences are negligible according to the Cliff's $\delta$ (which is below 0.15). Regarding the circumnuclear \ion{H}{ii} regions, the only effect we observe is that in galaxies with strong bars the N/O abundances are on average 0.05 dex lower than in galaxies with weak bars. Lastly, in the case of bar \ion{H}{ii} regions, some differences are found in $A_V$, but again these are negligible based on the Cliff's $\delta$ value.

Therefore, we conclude that, in general, the bar strength does not seem to significantly affect the properties of the \ion{H}{ii} regions, which present similar distributions of values independently of whether they are located in a galaxy with a weak or a strong bar. It is true, however, that the number of bar \ion{H}{ii} regions is significantly lower in galaxies with weaker bars compared to galaxies with stronger bars ($\sim 130$ against $400$, see Table~\ref{tab:sizes2}), although the coverage of the bar region is similar in both subsamples. The origin of this deficiency of SF ionised gas within the bar is unclear. \citet{diazgarcia2020} showed that SF along the bar (labelled as class C) is more typical of barred galaxies with high gravitational torques (i.e. strong bars), finding an increase in the frequency of this SF class with increasing $Q_{\rm bar}$. It is known that shear and shocks can limit SF, acting against the condensation of massive clouds \citep[e.g.][]{reynaud1998, zurita2004, seigar2005, kim2024}. The bar strength, as characterised in \citet{diazgarcia2020} and in our work, represents the gravitational bar torque measured by the tangential-to-radial force ratio ($F_T/⟨F_R⟩$). Tangential forces trace bar-induced gas motions, while radial forces control circular velocities in the inner parts, and thus the degree of shear. According to \citet{diazgarcia2020}, a larger shear is expected in weaker bars, and could explain the decline of SF along the bar found for these galaxies in our sample. However, shear depends on many other physical parameters than $⟨F_R⟩$ alone. For instance, the tangential forces remove angular momentum from the gas and produce the inflow, so the shear will depend on how much angular momentum the gas loses when it shocks in the leading edges of the bar. It will also depend on the relative motion of layers of gas which does not depend only on $⟨F_R⟩$ or angular momentum loss \citep[e.g.][]{emsellem2015}. The process is highly complex and further investigation is needed. Moreover, if we use $\epsilon_{\rm bar}$ to parametrise the bar strength, a similar number of \ion{H}{ii} regions are detected within the bar for galaxies with weak and strong bars. It is probable then that our result is biased by the low number of galaxies in the sample (17). This will have to be confirmed by future works extending the analysis to larger samples. 

\subsubsection{Galaxy mass}
The stellar mass of the galaxy is one of the most relevant parameters for evolutionary processes, as it set constraints on other important properties such as the metallicity or the SFR in a galaxy. Therefore, it is also pertinent to investigate whether the differences in the local properties of the \ion{H}{ii} regions depend on the host galaxy mass. For that, we repeat the analysis performed in Sec.~\ref{sec:results} but splitting now the sample in two subsets: low-mass galaxies, with $\log M_\star < 10.65 \,{\rm M_\odot}$, and massive barred systems, with $\log M_\star \geq 10.65 \,{\rm M_\odot}$. The limiting value has been set arbitrarily to have a similar number of objects in each subgroup, but we have checked that small changes in this limit do not significantly affect the results. We note that this analysis is limited by the small range of stellar masses covered by the sample (lower than 1 dex). Its novelty makes it worthwhile, but a better insight into the matter would be achievable in a future study based on a larger sample with a better coverage in terms of mass.

The corresponding p-values for the comparisons of the analysed property distributions of the different populations of \ion{H}{ii} regions for this segregation are shown in the third column of Table~\ref{tab:cliffs}. Again, the Cliff's $\delta$ values are also provided when the p-values are lower than 5\%. Interestingly, clear differences arise for all populations of \ion{H}{ii} regions. Overall, irrespectively of their location within the galaxy, \ion{H}{ii} regions in massive systems display larger O/H and N/O abundances, $A_V$, and $N_e$, although generally the differences are small ($\delta < 0.33$) but more significant in bar and circumnuclear regions. The differences in the chemical abundances are also confirmed at fixed $\Sigma_\star$ when inspecting the rMZR and the rMNOR, being on average of $\sim0.03$ dex for O/H and of $\sim0.07$ dex for N/O. These differences are small, and might be compatible with the absence of clear deviations found in the rMZR with the stellar mass based on lower spatial resolution IFU data \citep{sanchez2020}, which might dilute the small differences detected with the higher spatial resolution data used here. Regarding the larger values found for $A_V$ and $N_e$ in massive galaxies (especially for circumnuclear and bar \ion{H}{ii} regions), these are consistent with the increased amount of gas (and hence dust) associated with these systems \citep[e.g.][]{groves2023}.

In the case of $L({\rm H}\alpha)$, \ion{H}{ii} regions in the disc and the bar present lower values in more massive galaxies, but not those located in the innermost region (circumnuclear). These differences are small for the disc \ion{H}{ii} regions, and quite large for bar regions, with an overlap between both distributions of 73\% and 35\%, respectively. Disc and bar \ion{H}{ii} regions in low-mass galaxies are on average 2.1 and 27 times more luminous than in high-mass galaxies, respectively.

As in the case of the bar strength, we must be cautious since we have a limited number of galaxies in the sample (17) and the coverage in mass of the sample is small (lower than 1 dex). Further confirmation from studies based on larger samples is needed to confirm the observed trends.
   
\subsection{Effect of contamination from DIG}
When trying to determine the \ion{H}{ii} region properties, the contribution from the underlying DIG is often pointed out as a source of uncertainty \citep[e.g.][]{zhang2017}. Their classification itself is reported to be affected by applying a proper correction for the contribution of the DIG to the nebulae's spectra \citep{congiu2023}. For low-spatial resolution IFS surveys such as CALIFA \citep{sanchez2012a} or MaNGA \citep{bundy2015}, the individual spaxels cover kiloparsec-scale regions, and therefore, the emission is a mix of \ion{H}{ii} regions and DIG, making it difficult to disentangle. In the case of high-spatial data like TIMER, it is possible to spatially separate \ion{H}{ii} regions from DIG, which allows us to model the latter and properly decontaminate the contribution of this DIG from the emission of the \ion{H}{ii} regions \citep[e.g.][]{belfiore2022,lugoaranda2024}.
   
DIG not only has a significant impact on the H$\alpha$ emission observed in star-forming regions \citep[e.g.][]{zurita2000,oey2007}, but also on emission-line ratios \citep[e.g.][]{reynolds1998,sanders2017,zhang2017}. Therefore, the decontamination of DIG is critical for properly estimating properties relying on combinations of different line ratios, such as chemical abundances. Although line ratios such as [\ion{S}{ii}]$\lambda 6717, 6731$/H$\alpha$, [\ion{N}{ii}]$\lambda 6584$/H$\alpha$ are found to be enhanced in DIG with respect to \ion{H}{ii},regions, the behaviour of [\ion{O}{iii}]$\lambda 5007$/H$\beta$ is very diverse \citep[DIG can show either higher or lower values compared to \ion{H}{ii} regions;][]{zhang2017}, making the bias introduced by DIG uncertain and hard to predict a priori. 
   
{\ttfamily PYHIIEXTRACTOR} is not only an algorithm design to detect the \ion{H}{ii} regions, but it includes a procedure to model the DIG component and derive the emission of the \ion{H}{ii} regions alone with a proper decontamination from the DIG. This avoids biased measurements in the properties of the SF ionised gas, providing more accurate results. We now assess if the DIG contamination might have an impact on the uncovered effect of the bar on the properties of the \ion{H}{ii} regions. In the case of studies based on bright \ion{H}{ii} regions, it is generally expected that they suffer only negligible contamination, and their conclusions might not be much contingent upon a proper subtraction of the DIG contribution. However, this is especially relevant for the results based on lower spatial resolution like CALIFA or MaNGA. With this aim we follow an alternative approach in which the DIG contribution is not subtracted from the \ion{H}{ii} region emission, considering the integrated flux within these regions as emitted entirely by the ionizing nebulae.
   
\begin{figure}
   \centering
   \includegraphics[width=\hsize]{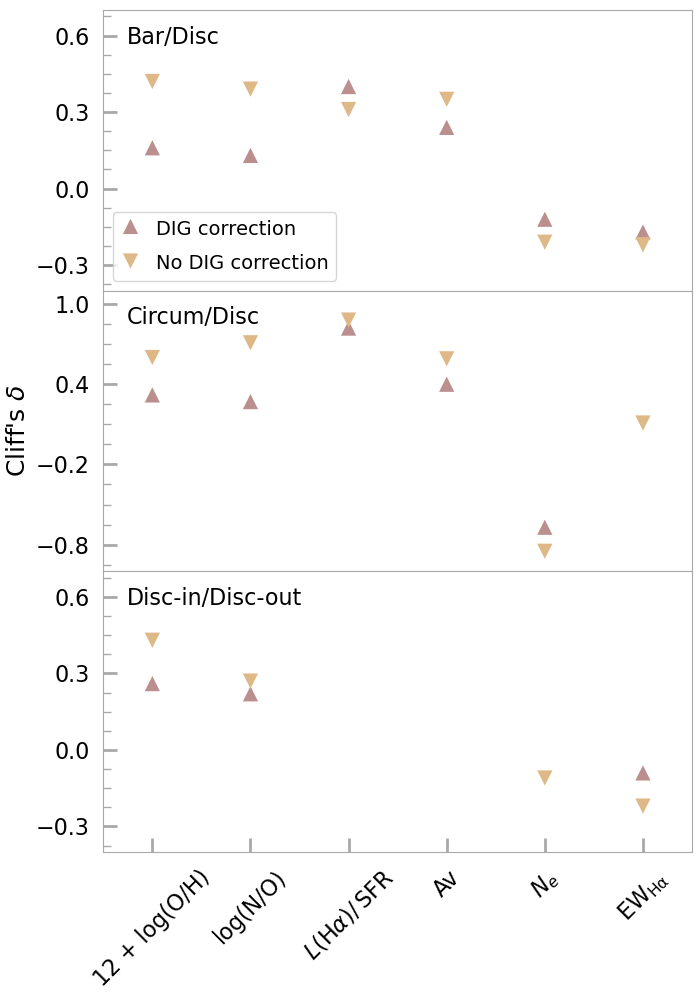}
      \caption{Cliff's $\delta$ values for the analysed properties of the compared subpopulations of \ion{H}{ii} regions with no decontamination from DIG emission (golden down triangles). The values for the analysis with DIG correction (described in Sec.~\ref{sec:results}) are shown for reference (brown up triangles). Only the cases where Mann-Whitney U p-values $<5\%$ are considered.
              }
         \label{fig:DIG_effect}
   \end{figure}

In Fig.~\ref{fig:DIG_effect} we show a comparison of the Cliff's $\delta$ values for the analysed properties of the explored subpopulations of \ion{H}{ii} regions when we do (brown up triangles) and do not (golden down triangles) perform the DIG decontamination. We only display the values in the cases where the distributions are statistically significantly different (as reflected by p-values below $5\%$ in the Mann-Whitney U test), when it is meaningful to quantify such differences using the Cliff's $\delta$ statistics. Overall, we can see that the trends are very similar in both cases. However, not correcting from the DIG emission generally results in higher $\delta$ values, revealing much larger differences than when we properly account for the DIG contribution. This is more evident when comparing the chemical abundances between disc and bar/circumnuclear \ion{H}{ii} regions. Other properties, such as the $L({\rm H}\alpha)$, the $N_e$ or the EW$_{\rm H\alpha}$ do not seem to be significantly affected by DIG. Within the disc, the effect of not correcting by DIG is smaller, although still present, particularly for the chemical abundances. In short, this analysis shows that the presence of the DIG introduces biases, proving the importance of a proper DIG correction in computing the \ion{H}{ii} region properties.

\subsection{Limitations of the study}\label{sec:caveats}

In Sec.~\ref{sec:abun}, we show that the \ion{H}{ii} regions located in the bar and the circumnuclear \ion{H}{ii} regions are more metal-rich than the regions in the disc. Within the disc, inner \ion{H}{ii} regions (inside the bar radius) have also higher oxygen abundances than the outer regions (outside the bar radius). A priori, one could expect that these differences might just be the result of the metallicity gradients observed in spiral galaxies due to the different ranges of galactocentric distances covered by each population of \ion{H}{ii} regions. Ideally, this could be solved by deriving the abundance gradients of the galaxy sample and analysing the residual abundances once this gradient is subtracted. However, our data only cover the central part of the galaxies, and therefore, a reliable abundance radial profile cannot be estimated. As a way to overcome this limitation, we have explored in Sec.~\ref{sec:rMZR} the rMZR, and whether the observed differences in the O/H abundances of the analysed populations of \ion{H}{ii} regions are only reflecting differences in $\Sigma_\star$. Although it might seem unrelated, the existence of the rMZR can explain the metallicity gradients simply considering the existence of an inverse radial gradient in $\Sigma_\star$ \citep[e.g.][]{gonzalezdelgado2014, sanchez2020}. Indeed, \citet{barreraballesteros2016} were able to reproduce the negative abundance gradient assuming a universal rMZR and a radial decline of $\Sigma_\star$. In this study we have assumed a connection between both observables and have inferred the effect of one of them by analysing the other. From this analysis we can deduce that in general, the differences in the oxygen abundance of the \ion{H}{ii} regions depending on their location within the galaxy are indeed a consequence of the underlying negative gradient, with the bar not playing a significant role in the chemical enrichment of the ionised gas. We have, however, some hints that the \ion{H}{ii} regions in the inner disc (inside the bar radius) are slightly more metal-rich than the \ion{H}{ii} regions in the outer disc (outside the bar radius) and the bar, not explained by the abundance gradient. It would be interesting to confirm these findings in fully-mapped spiral galaxies, where the abundance gradients can be derived and the abundance residuals analysed.

Due to the range of distances covered by the sample ($7-38.5$ Mpc), the mentioned limitation in the spatial coverage of the galaxies affects differently each of them. For the majority of the galaxies we are able to map between 4x4 and 7x7 kpc around the galaxy centre, although some galaxies are only covered in their central 2x2 kpc, and for others we reach up to 11 kpc. As a consequence, the bar region is not fully covered in many galaxies, hampering for instance the mapping of the bar ends in some cases. This part of the bar is specially interesting due to orbital crowding (overlap of different bar orbits and also between bar and spiral arms) and the intersection of gas on such orbits. This gas is then compressed and collapsed due to the slow down at the apocentres of the orbits, effects capable of enhancing the SFR and the star formation efficiency in these regions \citep[e.g.][]{kenney1991, lia1992, rodriguezfernandez2008, renaud2015, beslic2021}. Future studies fully mapping the bar region in a large sample of galaxies might enable to further separate the bar population between the \ion{H}{ii} regions distributed along the bar area and those found at its ends, allowing to investigate how the properties of the ionised gas might change also within this dynamical structure. We note that 5 galaxies ($30\%$) are also part of the PHANGS-MUSE sample and have available \ion{H}{ii} region catalogues  \citep{congiu2023, groves2023}. For consistency in addressing the analysis of the entire sample (same methodology to derive the \ion{H}{ii} region properties) we have preferred to restrict the study to the TIMER data, but a future study using the PHANGS-MUSE data would allow to extend the detection of the \ion{H}{ii} regions up to the outer discs and confirm the observed trends. For now, our study has been able to reveal significant differences in the properties of the ionised gas between the nuclear disc, the bar, and the disc components of the galaxies. It is true that the bulk of the population of disc \ion{H}{ii} regions belongs to just five galaxies in the sample (with three more presenting a couple of dozens disc \ion{H}{ii} regions, see Table~\ref{tab:sizes1} for a record of the number of \ion{H}{ii} regions in each analysed population associated with each galaxy in the sample). However, we believe these are representative of the entire population of disc \ion{H}{ii} regions, since the reported general trends are also observed in these individual galaxies.

\subsection{Comparison with previous studies}
One of the pioneering works addressing the properties of \ion{H}{ii} regions in bar environments was \citet[][hereafter MF99]{martin1999}. In particular, the authors studied the electron density, dust extinction, $\rm EW_{H\alpha}$, and O/H abundances for a sample of 50 bar and 40 disc \ion{H}{ii} regions from 10 barred galaxies\footnote{We note that there are no galaxies in common between our sample and that of \citet{martin1999}.}. They found that, compared to the disc, the SF ionised gas in the bar presents similar electron density, higher dust extinction and slightly lower $\rm EW_{H\alpha}$. Our results are all in agreement with those of MF99. Regarding the dust extinction, their differences are about 0.3 mag, which is an intermediate value between the differences in $A_V$ we find for the bar-disc comparison ($0.2$ mag) and the circumnuclear-disc comparison ($0.4$ mag). It is unclear whether MF99 include the circumnuclear \ion{H}{ii} regions in their sample of bar regions, since some of the galaxies present clear SF activity in this innermost part (like NGC~3504). The median values of $A_V$ differ slightly (1.7 and 1.5 for bar and disc \ion{H}{ii} regions in our work, respectively, against 1.3 and 1.0 mag in MF99), maybe due to the different laws used for reddening correction.

Just like MF99, we find equivalent $[\ion{S}{ii}]\lambda6717/[\ion{S}{ii}]\lambda6731$ line ratios for the bar and disc \ion{H}{ii} regions, reflecting comparable electron densities. The average values reported in MF99 are $1.33\pm0.02$ (bar regions) and $1.31\pm0.03$ (disc regions), very similar to ours ($1.38\pm0.05$ and $1.40\pm0.06$, respectively).

With respect to the $\rm EW_{H\alpha}$, MF99 reported slightly lower values for the bar regions, with a difference in the mean about a factor of two, indicating older ages for the \ion{H}{ii} regions in the bar, although the authors caution about the large uncertainties associated with their estimation introduced by the galactic continuum correction. \citet{briere2012}, who measured the age of the young populations contained in 51 giant \ion{H}{ii} regions in the spiral arms and bar of the galaxy NGC~5430 found no variation in age between the bar and disc \ion{H}{ii} regions. In our case, like MF99, we find slightly lower values of $\rm EW_{H\alpha}$ in the bar \ion{H}{ii} regions (about a factor of 1.3 in the median values). However, discrepancies in the methodology used to derive $\rm EW_{H\alpha}$ prevent us from doing a direct comparison of the $\rm EW_{H\alpha}$ values and their differences, since unlike the authors, we do not apply a correction for the contribution of the underlying stellar population when deriving $\rm EW_{H\alpha}$ (due to the difficulty in determining the fraction of the continuum emission not due to the ionizing star continuum). For this same reason, it is uncertain whether we can directly link our measurements of $\rm EW_{H\alpha}$ with the age of the \ion{H}{ii} regions, and therefore, with the results of \citet{briere2012}.

Finally, MF99 also explored the distribution of oxygen abundances of the \ion{H}{ii} regions within the bar, observing a very homogeneous distribution. Our finding of both circumnuclear and bar regions presenting similar values of O/H is in agreement with these results, as well as the slightly larger values found for the inner disc \ion{H}{ii} regions compared to the bar regions, indicative of an efficient mixing of the chemical composition within these central parts of the galaxy. \citet{briere2012} also investigated the oxygen abundances of the \ion{H}{ii} regions in NGC~5430, finding a very flat radial profile with no significant variation between the bar and spiral arm regions, consistent with the bar-induced mixing scenario. These results are not in disagreement with our reported O/H differences between the inner disc and bar \ion{H}{ii} regions. NGC~5430 presents a spiral structure with a very wide pitch angle, and since the arms emerge from the bar ends, most of the disc \ion{H}{ii} regions in the study should be located outside the bar radius, and would fall within the outer disc \ion{H}{ii} region category used in our study. We do not find any difference in the O/H abundances between the bar and outer disc \ion{H}{ii} regions (see top-middle panel of Fig.~\ref{fig:mzr}), which is consistent with the conclusions of \citet{briere2012}.

Around the same time, \citet{rozas1999} determined the luminosity function (LF) of the \ion{H}{ii} regions in the barred spiral NGC~7479 with separate analyses for the regions in the bar and in the disc. Using about 1000 \ion{H}{ii} regions, they showed that the LF of the bar is much less regular than the LF of the disc, with the bar having a higher number of very luminous regions, implying different SF conditions in the bar and the disc. Consistent with this picture, we find higher H$\alpha$ luminosities for the \ion{H}{ii} regions in the bar compared to disc regions, providing further evidence of the enhancement in SF driven by the bar.

Another interesting early work, this time focused on the properties of circumnuclear \ion{H}{ii} regions, was \citet[][hereafter K89]{kennicutt1989b}. Among several other properties, the authors analysed the H$\alpha$ luminosity, electron density, dust extinction, and $\rm EW_{H\alpha}$ for a compilation of (circum)nuclear and disc \ion{H}{ii} regions. They found that, compared to the disc, the SF ionised gas in the (circum)nuclear region presents similar luminosities and dust extinction, higher electron density, and lower $\rm EW_{H\alpha}$. Qualitatively, our results are in agreement with those of K89 regarding the electron density, but differ in the rest of their conclusions. Unlike K89, we do find significantly higher $A_V$ for the circumnuclear \ion{H}{ii} regions. The authors report a large dispersion in the values for these regions, with a few of them suffering heavier extinction than any of the disc \ion{H}{ii} regions, but with comparable median values for both populations. The low statistics in the circumnuclear subsample ($\sim 55$) in K89 compared to ours ($\sim 300$) might be hindering the detection of the observed differences. On the other hand, regarding $L({\rm H}\alpha)$, the fact that, unlike in our case, these are not corrected for extinction in K89 might be partially responsible of the found discrepancy, however, the large differences reported in our study (on average circumnuclear \ion{H}{ii} regions are 17 times more luminous in H$\alpha$ than disc regions) should be observable even in the absence of dust extinction correction. Another possible cause is the fact that in K89 the disc population is composed by \ion{H}{ii} regions outside the nuclear regions ($R > 1$ kpc), and the authors do not seem to make any distinction between those that are located in the disc and those that might belong to the bar region. The contamination of the disc population by \ion{H}{ii} regions in the bar can indeed decrease the differences in $L({\rm H}\alpha)$ since, as reported here, bar regions are also more luminous than disc regions, but again, it is unlikely that this factor can totally blur the large differences detected in $L({\rm H}\alpha)$ between the circumnuclear and the disc \ion{H}{ii} regions. Lastly, unlike K89, we do not find significantly lower $\rm EW_{H\alpha}$ in circumnuclear regions compared to disc regions, but similar distributions. However, as it happens with MF99, discrepancies in the methodology used to derive this parameter (we do not correct for the contribution of the underlying stellar population) prevent us from doing a direct comparison with K89. In any case, we are inclined to consider that the advances in the observational techniques since K89, that have helped to improve the analysis both in statistics and in resolution (spatial and spectral) of the data, have allowed us to reveal the existing differences in the properties of the circumnuclear \ion{H}{ii} regions that were hidden before.

Excluding these few works, that were focused on exploring the properties of the \ion{H}{ii} regions in barred galaxies, most of the studies addressing the impact of bars in the physical properties of the ionised gas rely on single-fibre apertures to compare the central properties of barred and unbarred systems. For these, we can only make a qualitative comparison of the results and their interpretation regarding the bar effect. For instance, our larger values of $L({\rm H}\alpha)$ in the bar and circumnuclear \ion{H}{ii} regions with respect to the disc regions are consistent with many studies unveiling the role of the bar driving the gas infall towards the central part of the galaxies to trigger SF \citep[e.g.][]{ellison2011, oh2012, florido2015, lin2017, lin2020, yu2022}. The finding of similar O/H abundances in bar and outer disc \ion{H}{ii} regions agrees also with previous studies that do not find differences between barred and unbarred systems \citep{cacho2014, florido2015}, but contradicts others \citep{ellison2011}. In this case, the source of the discrepancy could be the use of the [\ion{N}{ii}]/[\ion{O}{ii}] empirical calibration \citep{kewley2002} by \citeauthor{ellison2011}, which depends on the N/O abundance. Their reported larger O/H abundances in barred galaxies could instead be reflecting the larger N/O abundances found in this study for the circumnuclear \ion{H}{ii} regions \citep[as suggested by][]{florido2015}. Our results are also consistent with the larger dust content, electron density, and N/O in the centres of barred galaxies compared to their unbarred counterparts reported by \citet{florido2015}. The SDSS fibre size of 3'' used in the study of \citet{florido2015}, and the covered redshift range of the sample imply that they restrict to the central few kpc ($\sim0.5-2$ kpc), which would mainly map the ionised gas corresponding to the circumnuclear \ion{H}{ii} regions in our sample, where we find the most extreme values for these parameters and the largest differences compared with the disc regions. 

\section{Conclusions}\label{sec:concl}
In this work we investigate the properties of the ionised gas in nearby galaxies, exploring possible differences in the \ion{H}{ii} regions placed at different locations. We use IFS data for 17 galaxies from the TIMER project, with an average spatial resolution of $80-90$ pc. Applying {\ttfamily PYHIIEXTRACTOR}, we detect the clumpy ionised regions and decontaminate their emission from the underlying diffuse ionised gas. Based on the WHaD diagram, we select the \ion{H}{ii} regions, obtaining a total sample of 2200 regions. They are then separated between those that are located in the nuclear disc, the bar, and the galactic disc. The disc population is further divided in inner and outer regions depending on whether they are located inside or outside the bar radius.

The main findings of the study are summarised below:
\begin{enumerate}
\item \ion{H}{ii} regions in the bar are more luminous and have higher dust extinction and lower EW$_{\rm H\alpha}$ than in the disc. The higher values observed in the O/H and N/O abundances of bar \ion{H}{ii} regions are explained due to the higher stellar mass surface densities in accordance with the local mass-metallicity and mass-N/O relations. At fixed $\Sigma_\star$, bar and disc regions present a similar degree of chemical enrichment.\\[-0.2cm]

\item Circumnuclear \ion{H}{ii} regions present the most extreme values in the analysed properties, being the most luminous regions and those having the highest N/O abundances, dust extinctions, and electron densities. \\[-0.2cm]

\item Within the disc, the population of \ion{H}{ii} regions present very similar properties irrespectively of whether they are inside or outside the bar radius, except for the O/H and N/O abundances, for which slightly higher values are found within the bar radius and also compared to the bar regions.\\[-0.2cm]

\item Bar strength (measured with $Q_{\rm bar}$) does not affect, in general, the properties of the \ion{H}{ii} regions, as we observe the same distribution of values in the presence of a weak or a strong bar. There are however some exceptions, like the finding of 2.5 times more luminous \ion{H}{ii} regions in the disc of galaxies with stronger bars, that would need to be confirmed with larger samples. The number of bar \ion{H}{ii} regions is significantly lower in galaxies with weaker bars. This does not happen when we parametrise the bar strength via $\epsilon_{\rm bar}$, suggesting that it might be a sample size effect. \\[-0.2cm]

\item \ion{H}{ii} regions in massive systems display larger O/H and N/O abundances (at fixed $\Sigma_\star$), $A_V$ and $N_e$, and lower $L({\rm H}\alpha)$ than in low-mass galaxies. These differences are larger in the bar and circumnuclear populations of \ion{H}{ii} regions. \\[-0.2cm]

\item Not decontaminating the \ion{H}{ii} region emission from the DIG contribution falsely enhances the differences in the chemical abundances of the \ion{H}{ii} regions, whereas other properties such as the $L({\rm H}\alpha)$ or the electron density are mainly unaffected.
\end{enumerate}

Our results support the scenario in which the redistribution of matter induced by the bar increases the molecular gas concentration in the central parts, which leads to an enhancement in the dust content, $L({\rm H}\alpha)$ (linked to the mass of the ionising star cluster by previous studies), and electron density of the ionised gas in the bar. Since this gas is funnelled towards the galaxy centre, the most extreme values in these properties are found for the circumnuclear \ion{H}{ii} regions. In addition, as the bar evolves, the availability of gas in the disc diminishes, decreasing the rate at which the material is transported to the central parts, and therefore, at which the stars form in the bar. This is supported by the higher N/O values found in these inner regions. Finally, the gas transported by the bar from the outer to the inner parts of the galaxies, which is metal poor due to the existent negative metallicity gradients, travels and is mixed within the bar, diluting the chemical content of the \ion{H}{ii} regions in the bar as compared to the ionised gas in the disc at $r<R_{\rm bar}$.

The larger values found for $A_V$ and $N_e$ in massive galaxies (especially for circumnuclear and bar \ion{H}{ii} regions), are consistent with the increased amount of gas (and hence dust) associated with these systems. 

Although our results are robust based on the large statistics provided by the analysed number of \ion{H}{ii} regions, the sample is restricted to 17  galaxies with prominent bars, relatively massive ($M_\star > 10^{10.3} \,M_\odot$) and with a high presence of inner structures built by the bar (nuclear rings, discs, lenses, bars, etc.). Future studies based on a more representative and larger sample of barred galaxies with a more complete spatial coverage of their disc are needed in order to confirm our results and conclusions regarding the effect of the bar on galaxy evolution.

\section*{Acknowledgements}

We are grateful to the anonymous referee for the thorough reading of the manuscript and very helpful and constructive comments.

LSM acknowledges support from Juan de la Cierva fellowship (IJC2019-041527-I) and from project PID2020-114414GB-100, financed by MCIN/AEI/10.13039/501100011033. JFB acknowledges support from the PID2022-140869NB-I00 grant from the Spanish Ministry of Science and Innovation. TK acknowledges support from the Basic Science Research Program through the National Research Foundation of Korea (NRF) funded by the Ministry of Education (No. 2022R1A4A3031306, No. RS-2023-00240212). AdLC  acknowledges financial support from the Spanish Ministry of Science and Innovation (MICINN) through RYC2022-035838-I and PID2021-128131NB-I00 (CoBEARD project). AZLA acknowledges support from the Universidad Nacional Autónoma de México Postdoctoral Program (POSDOC).

This research was supported by the Munich Institute for Astro-, Particle and BioPhysics (MIAPbP), which is funded by the Deutsche Forschungsgemeinschaft (DFG, German Research Foundation) under Germany´s Excellence Strategy – EXC-2094 – 390783311.

Funding for open access charge: Universidad de Granada / CBUA.

\section*{Data Availability}

The raw and reduced MUSE data used in this work are publicly
available at the ESO Science Archive Facility. The catalogue of \ion{H}{ii} regions is published along with this paper and available in the online supplementary material of the journal.



\bibliographystyle{mnras}
\bibliography{bibliography} 




\appendix

\section{Atlas of the galaxy sample}\label{sec:ap1}
Figure~\ref{fig:atlas} shows H$\alpha$ images (in logarithmic scale) of the galaxies in the sample  (except for NGC~4981 that can be found in Fig.~\ref{fig:HIIregions}). The dashed orange circles and solid red ellipses delimit the nuclear disc and bar regions, respectively. We do not cover the full extension of the bar for some galaxies, but within the mapped area, we can already see a variety of behaviours regarding the location of the \ion{H}{ii} regions: 8 galaxies display extensive SF along the bar region, whereas 9 of them only exhibit localised SF arranged in an inner ring or placed at the ends of the bar.

\begin{figure*}
   \centering
   \includegraphics[width=\hsize]{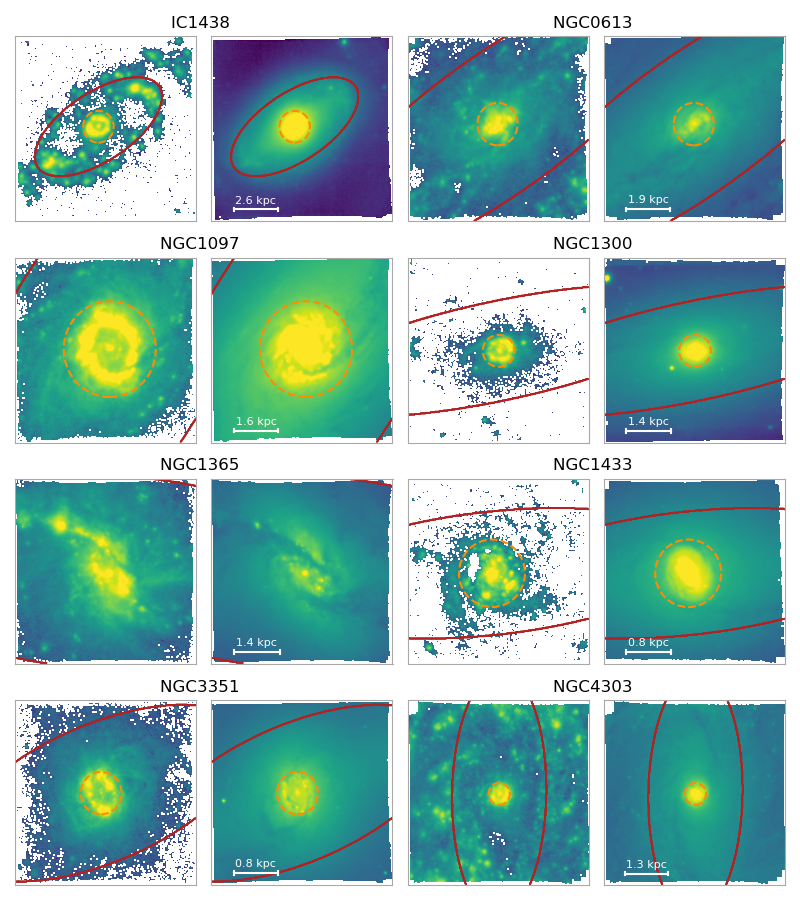}
      \caption{H$\alpha$ ({\it left}) and continuum ({\it right}) images (in logarithmic scale) of the galaxies in the sample. The dashed orange circles and solid red ellipses delimit the nuclear disc and bar regions, respectively.}
         \label{fig:atlas}
   \end{figure*}

\begin{figure*}
   \centering
   \includegraphics[width=\hsize]{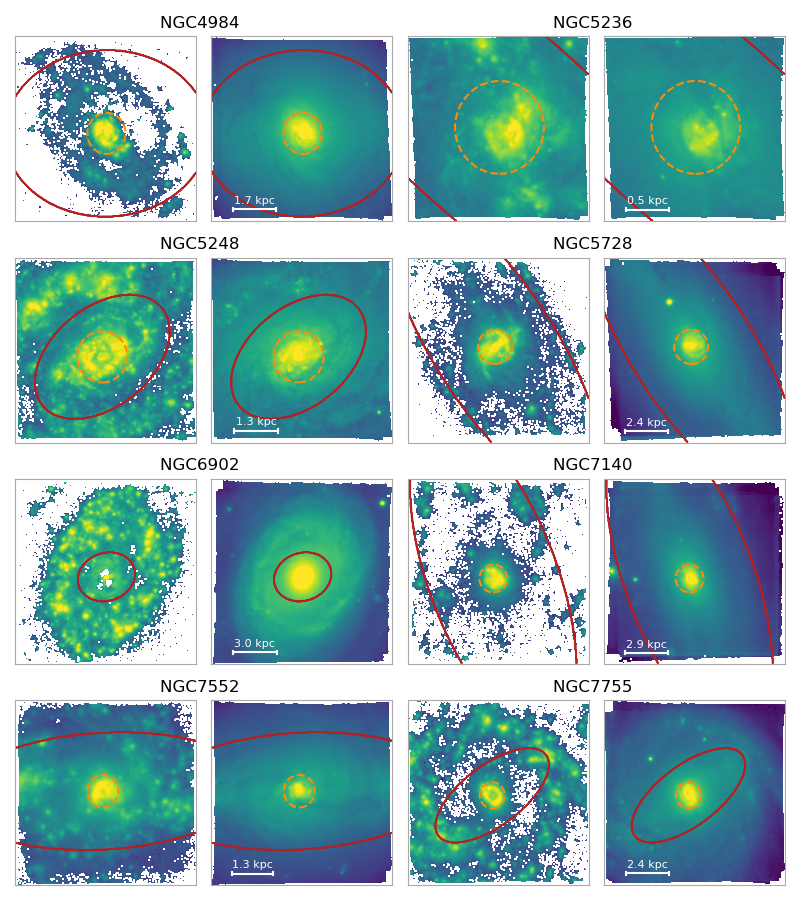}
      \contcaption{}
   \end{figure*}

\section{Catalogue of \ion{H}{ii} regions}\label{sec:ap3}
In this Appendix we provide details on the catalogue of \ion{H}{ii} regions that will be released along with the paper. This information has been divided in Tables~\ref{tab:table1_app}-\ref{tab:table4_app} for displaying purposes but will be published as a single catalogue in its entirety in the machine readable format. Only a portion containing the first 10 rows is shown here for guidance regarding its form and content. In this catalogue we provide information for the \ion{H}{ii} regions on the flux of the main strong emission lines (Tables~\ref{tab:table1_app} and \ref{tab:table2_app}), and some physical properties (Tables~\ref{tab:table3_app} and \ref{tab:table4_app}). In addition to the \ion{H}{ii} region properties analysed in Sec.~\ref{sec:results}, we also include the equatorial coordinates, the radius of the regions (defined as $\sqrt{2} \sigma$, where $\sigma$ is the dispersion of the Gaussian function used to model the \ion{H}{ii} regions; see \citealt[][for details]{lugoaranda2022}), the environment in which they belong (circumnuclear region, bar, inner disc, and outer disc; see Sec.~\ref{sec:population} for details), the velocity dispersion of H$\alpha$ (used to classify the regions as star forming; see Sec.~\ref{sec:hiiregions}) and the stellar mass surface density (used for the analysis of the resolved mass-metallicity relation; see Sec.~\ref{sec:rMZR}). All the measurements are provided together with their corresponding uncertainties.

\begin{table*}
\caption{Emission line fluxes up to [\ion{N}{ii}]$\lambda6548$ contained in the published catalogue of \ion{H}{ii} regions. Fluxes are all in units of $10^{-20}$ erg/s/cm$^2$.}
\label{tab:table1_app}
\begin{tabular}{lcccccccc}
\hline
RegionID & H$\beta$ & e\_H$\beta$ & [\ion{O}{iii}]$\lambda4959$ & e\_[\ion{O}{iii}]$\lambda4959$ & [\ion{O}{iii}]$\lambda5007$ & e\_[\ion{O}{iii}]$\lambda5007$ & [\ion{N}{ii}]$\lambda6548$ & e\_[\ion{N}{ii}]$\lambda6548$ \\
\hline
IC1438-01  &  126013.99  &  443.34  &  4413.16  &  99.50  &  13370.59  &  301.52  &  73808.33  &  117.44  \\ 
IC1438-02  &  84729.04  &  460.12  &  2883.53  &  112.27  &  8721.24  &  340.22  &  49171.57  &  114.22  \\ 
IC1438-03  &  84577.46  &  351.20  &  9694.31  &  65.33  &  29376.69  &  197.97  &  45635.33  &  76.34  \\ 
IC1438-04  &  79825.77  &  383.65  &  3548.17  &  89.88  &  10752.02  &  272.35  &  48447.54  &  110.24  \\ 
IC1438-05  &  51705.71  &  426.98  &  2695.68  &  112.05  &  8151.70  &  339.55  &  34655.89  &  117.29  \\ 
IC1438-06  &  41712.97  &  438.27  &  3923.87  &  115.73  &  11890.51  &  350.69  &  33254.45  &  131.45  \\ 
IC1438-07  &  41846.92  &  257.98  &  3527.38  &  54.75  &  10680.20  &  165.90  &  26922.47  &  64.33  \\ 
IC1438-08  &  41018.17  &  265.80  &  3278.24  &  58.54  &  9919.66  &  177.40  &  26544.92  &  63.54  \\ 
IC1438-09  &  35736.56  &  355.70  &  2990.10  &  93.15  &  9060.90  &  282.27  &  26594.56  &  100.37  \\ 
IC1438-10  &  34216.07  &  256.50  &  3976.08  &  60.89  &  12034.80  &  184.53  &  20966.98  &  55.25  \\ 
\hline
\end{tabular}
\end{table*}

\begin{table*}
\caption{Emission line fluxes from H$\alpha$ on contained in the published catalogue of \ion{H}{ii} regions. Fluxes are all in units of $10^{-20}$ erg/s/cm$^2$.}
\label{tab:table2_app}
\begin{tabular}{lcccccccc}
\hline
RegionID & H$\alpha$ & e\_H$\alpha$ & [\ion{N}{ii}]$\lambda6583$ & e\_[\ion{N}{ii}]$\lambda6583$ & [\ion{S}{ii}]$\lambda6717$ & e\_[\ion{S}{ii}]$\lambda6717$ & [\ion{S}{ii}]$\lambda6731$ & e\_[\ion{S}{ii}]$\lambda6731$\\
\hline
IC1438-01  &  586168.50  &  645.01  &  223661.61  &  355.87  &  84651.88  &  265.88  &  62924.11  &  244.42  \\ 
IC1438-02  &  389760.19  &  577.88  &  149004.75  &  346.12  &  53332.78  &  296.98  &  39134.69  &  268.17  \\ 
IC1438-03  &  360755.66  &  439.54  &  138288.89  &  231.34  &  43772.22  &  164.93  &  31569.55  &  153.97  \\ 
IC1438-04  &  373055.94  &  541.66  &  146810.71  &  334.07  &  59847.25  &  258.28  &  43921.69  &  231.66  \\ 
IC1438-05  &  250178.83  &  559.82  &  105017.86  &  355.41  &  37971.93  &  301.22  &  28688.52  &  276.70  \\ 
IC1438-06  &  199513.30  &  565.54  &  100771.04  &  398.34  &  27510.42  &  279.01  &  21693.74  &  283.03  \\ 
IC1438-07  &  186200.59  &  324.43  &  81624.48  &  194.94  &  30318.62  &  159.00  &  22052.69  &  143.47  \\ 
IC1438-08  &  184447.09  &  311.94  &  80496.05  &  192.54  &  36165.89  &  159.10  &  26614.56  &  137.18  \\ 
IC1438-09  &  178631.15  &  442.85  &  80589.59  &  304.14  &  32931.91  &  255.90  &  24880.04  &  243.82  \\ 
IC1438-10  &  163014.43  &  286.61  &  63597.99  &  167.42  &  22993.89  &  126.91  &  16431.88  &  116.19  \\ 
\hline
\end{tabular}
\end{table*}

\begin{table*}
\caption{First set of properties contained in the published catalogue of \ion{H}{ii} regions.}
\label{tab:table3_app}
\begin{tabular}{lccccccccc}
\hline
RegionID & RA$_{\rm reg}$ & Dec$_{\rm reg}$ & Radius & Environment & $\sigma({\rm H}\alpha)$ & e\_$\sigma({\rm H}\alpha)$ & EW$_{\rm H\alpha}$ & e\_EW$_{\rm H\alpha}$ & $\log \Sigma_\star$ \\
 & {\footnotesize [$\degr$]} & {\footnotesize [$\degr$]} & {\footnotesize [pix]} &  & {\footnotesize [km s$^{-1}$]} & {\footnotesize [km s$^{-1}$]} & {\footnotesize [\AA]} & {\footnotesize [\AA]} & {\footnotesize [$\rm M_{\sun} \, pc^{-2}$]} \\
\hline
IC1438-01  &  334.1230  &  -21.4311  &  3.1  &  Circum  &  24.35  &  4.50  &  60.12  &  0.07  &  3.124  \\ 
IC1438-02  &  334.1227  &  -21.4313  &  3.1  &  Circum  &  24.33  &  6.14  &  31.64  &  0.05  &  3.229  \\ 
IC1438-03  &  334.1182  &  -21.4271  &  3.1  &  Bar  &  24.47  &  5.16  &  146.94  &  0.25  &  2.562  \\ 
IC1438-04  &  334.1231  &  -21.4308  &  3.1  &  Circum  &  25.52  &  6.12  &  34.40  &  0.05  &  3.045  \\ 
IC1438-05  &  334.1224  &  -21.4314  &  3.1  &  Circum  &  26.33  &  9.05  &  15.78  &  0.04  &  3.295  \\ 
IC1438-06  &  334.1220  &  -21.4300  &  3.1  &  Circum  &  26.29  &  11.14  &  10.09  &  0.03  &  3.418  \\ 
IC1438-07  &  334.1272  &  -21.4339  &  3.1  &  Bar  &  21.22  &  7.95  &  166.89  &  0.52  &  2.479  \\ 
IC1438-08  &  334.1273  &  -21.4343  &  3.1  &  Bar  &  20.80  &  7.48  &  186.93  &  0.59  &  2.397  \\ 
IC1438-09  &  334.1212  &  -21.4310  &  3.1  &  Circum  &  29.19  &  10.65  &  15.09  &  0.04  &  3.969  \\ 
IC1438-10  &  334.1276  &  -21.4345  &  3.1  &  Bar  &  16.02  &  10.60  &  146.32  &  0.41  &  2.294  \\ 
\hline
\end{tabular}
\end{table*}

\begin{table*}
\caption{Second set of properties contained in the published catalogue of \ion{H}{ii} regions.}
\label{tab:table4_app}
\begin{tabular}{lcccccccc}
\hline
RegionID & $12+\log \rm O/H$ & e\_O/H & $\log {\rm N/O}$ & e\_$\log {\rm N/O}$ & $A_V$ & e\_$A_V$ & $\log L({\rm H\alpha})$ & e\_$\log L({\rm H\alpha})$ \\
 & {\footnotesize [dex]} & {\footnotesize [dex]} & {\footnotesize [dex]} & {\footnotesize [dex]} & {\footnotesize [mag]} & {\footnotesize [mag]} & {\footnotesize [erg s$^{-1}$]} & {\footnotesize [erg s$^{-1}$]} \\
\hline
IC1438-01  &  8.586  &  0.001  &  -0.908  &  0.001  &  1.525  &  0.012  &  39.404  &  0.004  \\ 
IC1438-02  &  8.597  &  0.001  &  -0.886  &  0.001  &  1.490  &  0.018  &  39.215  &  0.006  \\ 
IC1438-03  &  8.611  &  0.001  &  -0.840  &  0.001  &  1.253  &  0.014  &  39.104  &  0.004  \\ 
IC1438-04  &  8.579  &  0.001  &  -0.933  &  0.001  &  1.539  &  0.016  &  39.212  &  0.005  \\ 
IC1438-05  &  8.606  &  0.001  &  -0.893  &  0.001  &  1.648  &  0.027  &  39.075  &  0.009  \\ 
IC1438-06  &  8.671  &  0.001  &  -0.798  &  0.001  &  1.612  &  0.034  &  38.964  &  0.011  \\ 
IC1438-07  &  8.612  &  0.001  &  -0.899  &  0.001  &  1.385  &  0.020  &  38.860  &  0.007  \\ 
IC1438-08  &  8.582  &  0.001  &  -0.970  &  0.001  &  1.419  &  0.021  &  38.867  &  0.007  \\ 
IC1438-09  &  8.602  &  0.001  &  -0.937  &  0.001  &  1.750  &  0.032  &  38.962  &  0.011  \\ 
IC1438-10  &  8.598  &  0.001  &  -0.885  &  0.001  &  1.600  &  0.024  &  38.873  &  0.008  \\  
\hline
\end{tabular}
\end{table*}

\section{Statistics on \ion{H}{ii} region sample sizes}\label{sec:ap2}
In this Appendix we provide information on the numbers of analysed \ion{H}{ii} regions for the different subsamples and segregations performed along the study. In particular, Table~\ref{tab:sizes1} shows the number of detected \ion{H}{ii} regions for each galaxy in the sample separated in the different populations according to their location within the galaxy: the nuclear disc, the bar, and the disc. The numbers for the subdivision of the disc population in inner and outer regions is also indicated within brackets. 

Table~\ref{tab:sizes2} compiles the number of \ion{H}{ii} regions comprising the different populations (disc, bar, circumnuclear) with measurements of each of the analysed properties. We also include the numbers for the two compared subgroups when segregating the galaxies according to the bar strength or the stellar mass.

\begin{table}
 \caption[]{\label{tab:sizes1} Number of detected \ion{H}{ii} regions for each galaxy in the sample separated according to their location (circumnuclear, bar, disc).}
\begin{tabular}{l@{\hskip 0.7cm}c@{\hskip 0.7cm}c@{\hskip 0.7cm}c@{\hskip 0.7cm}c}
 \hline \hline\\[-0.3cm]
 Galaxy & Total & Circum. & Bar & Disc\\ \hline \\[-0.2cm]
IC~1438 & $72$ & $11$ & $38$ & $23$ \\
 & & & & ($14/9$) \\[0.2cm]
NGC~0613 & $77$ & $4$ & $50$ & $23$ \\
 & & & & ($23/0$) \\[0.2cm]
NGC~1097 & $67$ & $49$ & $16$ & $2$ \\
 & & & & ($2/0$) \\[0.2cm]
NGC~1300 & $22$ & $20$ & $2$ & $0$ \\[0.2cm]
NGC~1365 & $66$ & $0$ & $66$ & $0$ \\[0.2cm]
NGC~1433 & $21$ & $17$ & $4$ & $0$ \\[0.2cm]
NGC~3351 & $39$ & $39$ & $0$ & $0$ \\[0.2cm]
NGC~4303 & $278$ & $16$ & $105$ & $157$ \\
 & & & & ($146/11$) \\[0.2cm]
NGC~4981 & $340$ & $12$ & $10$ & $318$ \\
 & & & & ($6/312$) \\[0.2cm]
NGC~4984 & $48$ & $26$ & $18$ & $4$ \\
 & & & & ($4/0$) \\[0.2cm]
NGC~5236 & $92$ & $36$ & $54$ & $2$ \\
 & & & & ($2/0$) \\[0.2cm]
NGC~5248 & $299$ & $59$ & $85$ & $155$ \\
 & & & & ($19/136$) \\[0.2cm]
NGC~5728 & $62$ & $14$ & $41$ & $7$ \\
 & & & & ($7/0$) \\[0.2cm]
NGC~6902 & $213$ & $0$ & $3$ & $210$ \\
 & & & & ($0/210$) \\[0.2cm]
NGC~7140 & $32$ & $11$ & $17$ & $4$ \\
 & & & & ($4/0$) \\[0.2cm]
NGC~7552 & $123$ & $1$ & $104$ & $18$ \\
 & & & & ($18/0$) \\[0.2cm]
NGC~7755 & $349$ & $16$ & $48$ & $285$ \\
 & & & & ($15/270$) \\[0.2cm]
\hline
\multicolumn{5}{l}{Numbers on inner and outer \ion{H}{ii} regions within the disc population are}\\
\multicolumn{5}{l}{also indicated within brackets.}\\
\end{tabular}
\end{table}

\begin{table}
 \caption[]{\label{tab:sizes2} Number of \ion{H}{ii} regions with available values for each of the analysed properties and comprising the different subpopulations (disc, bar, circumnuclear) for each explored subgroup of galaxies.}
\begin{tabular}{l@{\hskip 0.7cm}c@{\hskip 0.8cm}c@{\hskip 0.6cm}c}
 \hline \hline\\[-0.1cm]
  & {\bf Total} & {\bf Bar strength} & {\bf Galaxy mass} \\[0.2cm] \hline \\[-0.2cm]
{\bf 12+log(O/H)} &  &  & \\
Disc & $769$ & $484/285$ & $408/361$ \\
Bar & $489$ & $114/375$ & $184/305$ \\
Circum & $306$ & $189/117$ & $136/170$   \\[0.2cm]
{\bf log(N/O)} &  &  & \\
Disc & $769$ & $484/285$ & $408/361$ \\
Bar & $489$ & $114/375$ & $184/305$\\
Circum & $306$ & $189/117$ & $136/170$ \\[0.2cm]
{\bf $\mathbf{L(H\alpha})$} &  &  &  \\
Disc & $796$ & $500/296$ & $427/369$ \\
Bar & $509$ & $124/385$ & $194/315$ \\
Circum & $306$ & $189/117$ & $136/170$ \\[0.2cm]
$\mathbf{A_V}$ &  &  & \\
Disc & $796$ & $500/296$ & $427/369$ \\
Bar & $509$ & $124/385$ & $194/315$ \\
Circum & $306$ & $189/117$ & $136/170$ \\[0.2cm]
{\bf $\mathbf{N_e}$} &  &  & \\
Disc & $913$ & $555/358$ & $458/455$ \\
Bar & $555$ & $135/420$ & $182/373$ \\
Circum & $330$ & $195/135$ & $141/189$ \\[0.2cm]
{\bf $\mathbf{ EW_{H\alpha}}$} &  &  & \\
Disc & $1208$ & $712/496$ & $648/560$ \\
Bar & $661$ & $170/491$ & $224/437$  \\
Circum & $331$ & $196/135$ & $142/189$ \\
\hline
\multicolumn{4}{l}{Galaxies have been grouped using the following criteria: (a) Bar}\\
\multicolumn{4}{l}{strength: galaxies with weak ($Q_{\rm bar}<0.3$) vs. strong ($Q_{\rm bar}\geq0.3$)}\\
\multicolumn{4}{l}{ bars; (b) galaxy mass: low- ($\log M_\star <10.65 \,{\rm M_\odot}$) vs. high-mass}\\
\multicolumn{4}{l}{($\log M_\star \geq10.65 \,{\rm M_\odot}$) galaxies.}\\
\end{tabular}
\end{table}


\bsp	
\label{lastpage}
\end{document}